%% file: arxiv.tex
\documentclass[nojss]{jss}

\usepackage{multirow}
\usepackage{rotating}
\usepackage{amsmath,amssymb}
\usepackage{bm}
\usepackage{enumerate}

\title{Revisiting Group Differences in High-Dimensional Choices:
  Method and Application to Congressional Speech}

\Shorttitle{Revisiting Group Differences in High-Dimensional Choices}

\author{Paul Hofmarcher\\Paris Lodron Universit\"at Salzburg \And
  Jan V{\'a}vra\\Paris Lodron Universit\"at Salzburg and\\Wirtschaftsuniversit\"at Wien \AND
  Sourav Adhikari\\Wirtschaftsuniversit\"at Wien \And Bettina
  Gr\"un\\Wirtschaftsuniversit\"at Wien}
\Plainauthor{Paul Hofmarcher, Jan V{\'a}vra, Sourav Adhikari, Bettina Gr\"un}

\Address{
  Paul Hofmarcher\\
  Department of Economics\\
  Paris Lodron University of
  Salzburg\\
  Mönchsberg 2a\\
  5020 Salzburg, Austria\\
  E-mail: \email{paul.hofmarcher@plus.ac.at}
}

\Abstract{
    Gentzkow, Shapiro and Taddy, Econometrica Vol 87, No 4, 2019
    (henceforth GST) use a supervised text-based regression model to
    assess changes in partisanship in U.S.\ congressional speech over
    time. Their estimates imply that partisanship is far greater in
    recent years than in the past, and that it increased sharply in
    the early 1990s. The paper at hand provides a replication in the
    wide sense of GST by complementing their analysis in three
    ways. First, we propose an alternative unsupervised language
    model, which combines ideas of topic models and ideal point
    models, to analyze the change in partisanship over time.  We apply
    this model to the Senate speech data used in GST ranging from
    1981--2017. Using our model we replicate their results on the
    specific evolution of partisanship. Second, our model provides
    additional insights such as the data-driven estimation of
    evolvement of topical contents over time.  Third, we identify key
    phrases of partisanship on topic level.
}


\begin{document}

\section{Introduction}\label{sec:intro}
Estimating political positions -- so called ideal points -- of
political key actors has a long tradition in political science
\citep[see, e.g.,][]{Poole:1985,McCarty+Poole+Rosenthal:1997}.
\citet{Taddy_etal_2019} (henceforth GST) contribute to this literature
on an aggregate level by proposing a text-based regression model to
study to what extent partisanship between political parties, estimated
from differences in word usage in speeches, evolves over time. They
use transcripts of the speeches given in the U.S.\ Congress between
1873 and 2016 (Congress Sessions 43 to 114). Their main result is that
partisanship in Congress speeches is a relatively recent phenomenon.
Its inflection point coincides precisely with innovations in political
persuasion in the 1990s formulated under a platform called the
\emph{Contract with America}.  Methodologically GST build on
\cite{Taddy:2013} to learn a predictive model for party given word
usage. They also investigate differences in partisanship according to
22 topics which they identified manually based on domain knowledge by
assigning key terms to characterize them.

The paper at hand presents the results obtained with an alternative
statistical model to assess the changes in partisanship over time. Our
model is a time-varying (TV) extension of the text-based ideal point
model \citep[TBIP;][]{Vafa_etal_2020} (henceforth TV-TBIP). TV-TBIP
combines the class of topic models \citep[see][]{blei_etal_2003} and
text-based ideal point models \citep[see,
e.g.,][]{Slapin_2008,wordshoal_2016}. In contrast to the
regression-based approach pursued in GST, TV-TBIP is an unsupervised
modeling approach where topics are extracted in a data-driven way
allowing their evolvement over time.

We apply our model to speeches given in the U.S.\ Senate in sessions
97--114, i.e., between 1981 and 2017, making use of the speech text
data provided as supplementary material by GST\footnote{GST use
  hein-daily and hein-bound transcripts (see
  \url{https://data.stanford.edu/congress_text}), where hein-bound
  covers the time period 1873--2011 and hein-daily 1981--2017. Given
  the focus on partisanship during the 1990s and later in GST, we make
  only use of the hein-daily speeches covering 1981--2017 in our
  analysis.}. Our model estimates for each speaker a session-specific
ideological position on a one-dimensional latent scale. This latent
scale is shown to primarily capture the party differences. Ideal
points capturing the ideological positions thus allow obtaining an
aggregate measure of partisanship for each session. In line with GST,
we find that Congress Session~105 is the first session where a clear
separation between the party-specific distributions of the ideal
points is discernible.  In contrast to GST, our model provides
partisanship estimates on the speaker level and these individual
ideological positions allow to infer which speakers are at the
extremes of the latent scale or rather centrally located and to track
changes of their ideological positions over time.  TV-TBIP extends the
approach taken in GST by estimating the topics in a data-driven
way. In addition, TV-TBIP also allows term compositions of topics to
change over time to account for newly emerging or subduing sub-themes
within a topic. This is particularly important when analyzing more than
40 years of speeches given in the U.S.\ Senate.

The paper is organized as follows: Section~\ref{sec:TBIP} introduces
the time-varying text-based ideal point
(TV-TBIP). Section~\ref{sec:results} discusses the results and
insights gained from fitting the TV-TBIP model and contrasts to them
to those in GST.  Finally, Section~\ref{sec:conc} concludes.

\section{The Time-Varying Text-Based Ideal Point Model}\label{sec:TBIP}
The TBIP model proposed in \citet{Vafa_etal_2020} assumes that
documents are composed of a fixed set of topics (latent themes) where
ideological differences of the authors influence the term compositions
of topics. Considering a corpus of speeches covering an extensive time
period of about 40 years, we extend the TBIP model to a time-varying
version which allows the term compositions of the topics as well as
the individual ideological positions to adapt over time.

\subsection{Model Specification}\label{sec:model-specification}

The available data for each session or time point $t$ consist of the
text of the speeches given as well as the information on the speaker
giving the speech. Based on the bag-of-words assumption, the speech
data are summarized in a document-term matrix $\bm{c}^t$. This matrix
contains the frequency counts of the terms in each speech at time
point $t$ with the number of rows corresponding to the number of
speeches and the number of columns to the vocabulary size. The speaker
information is combined in the vector $\bm{s}_t$ with length equal to
the number of speeches.

The parameters and latent variables inferred in the TV-TBIP model for
each time point $t$ consist of document-specific topic prevalences
$\bm{\theta}^t$, topic-specific term prevalences for neutral speakers
$\bm{\beta}^t$, topic-specific polarity scores for the terms
$\bm{\eta}^t$ and individual ideal points (positions) of the speakers
$\bm{x}^t$.  The matrix of topic distributions at time point $t$,
$\bm{\theta}^t$, has dimension number of speeches times number of
topics. The matrix $\bm{\beta}^t$ is of dimension number of topics
times the size of the vocabulary. $\bm{\eta}^t$ has the same dimension
as $\bm{\beta}^t$ and the vector of ideal points $\bm{x}^{t}$ has the
length of the number of speakers at time $t$.

The TV-TBIP model assumes that the observed frequencies of the terms
in the speeches are generated independently from a Poisson
distribution, i.e., the frequency count $c^t_{iv}$ at time point $t$
for speech $i$ and term $v$ is generated by:
\begin{equation}
  c^t_{iv} \sim \text{Pois}(\sum_{k=1}^K \theta^t_{ik} \beta^t_{kv} \exp \{x^t_{s_i} \eta^t_{kv}\}).
\label{form:TBIP}
\end{equation}
The Poisson rates are derived as linear combinations of contributions
from the $K$ topics where $\theta^t_{ik}$ is the topic prevalence of
topic $k$ in speech $i$ at time point $t$ and $\beta^t_{kv}$ is the
term prevalence of term $v$ in topic $k$ at time point $t$ of a
neutral speaker, i.e., these term prevalences characterize ``neutral
topics''.

For each speaker $s$, the ideological position is encapsulated in the
ideal point $x^t_{s}$. For a neutral speaker with $x^t_{s}=0$ the
model reduces to the simple Poisson factorization topic model
\citep{John:2004}. With ideological positions departing from the
neutral one, the Poisson rates of the terms for each of the topics
also differ from the neutral one. If the ideal point $x^t_{s}$ and the
prevalence modification $\eta^t_{kv}$ at time point $t$ for term $v$
and topic $k$ have the same sign, a speaker will more often use this
term when talking about this topic compared to a neutral speaker. If
they have opposite signs, usage of that term for this topic is
decreased compared to someone who is neutral. The TV-TBIP model
assumes that the polarity of a speaker captured by the ideal points
influences their word choice for a specific topic, i.e., the term
prevalences of a specific topic differ between a negative, a neutral
and a positive speaker. By contrast, the model assumes that ideal
points do not impact on the topic prevalences, i.e., a negative, a
neutral and a positive speaker talk to a similar extent about each of
the topics.

\subsection{Model Estimation}
We estimate the model within a Bayesian framework. The parameters and
latent variables are assumed to be independent and identically
distributed with the same prior settings used for each time point,
speech, term and topic as well as speaker.
The priors for the topic and term prevalences $\theta^t_{ik}$ and
$\beta^t_{kv}$ are assumed to follow independent Gamma distributions
with potentially different hyperparameters for $\theta^t_{ik}$ and
$\beta^t_{kv}$, but identical Gamma distributions for all sessions
$t$, speeches $i$, terms $v$ and topics $k$:
\begin{align*}
\theta^t_{ik} &\sim \text{Gamma}(\alpha_1, \alpha_2),&
\beta^t_{kv} &\sim \text{Gamma}(\gamma_1, \gamma_2).
\end{align*}
Careful selection of the parameters of the Gamma distributions is
required in order to induce a suitable sparsity in the topic
prevalences of the speeches and in the term prevalences of the topics.
Sparsity implies that each speech consists of a small number of topics
and that each topic is characterized by a few terms with high
prevalence values. The choice of these hyperparameters impacts on the
specific solution obtained.  Here we follow \cite{Vafa_etal_2020} and
set $\alpha_1=\alpha_2=0.3$ and $\gamma_1=\gamma_2=0.3$. To facilitate
interpretability of the topics obtained, such a sparsity
characteristic is desirable. In addition, this alleviates the issue of
non-identifiability which topic models are known to suffer from and
which we would expect to also be an issue for TV-TBIP. For
non-identifiable topic models, \citet{mrlata2020} highlight the
importance of selecting suitable priors which support to prefer a
specific parameterization among those inducing the same likelihood.

The topic polarity scores and ideal points $\eta^t_{kv}$ and $x^t_{s}$
can take arbitrary values, negative as well as positive. To reflect
this unrestricted support, standard normal distributions are imposed
as their priors. This means that for all sessions $t$, terms $v$,
topics $k$, and speakers $s$, we assume independent distributions
given by:
\begin{align*}
\eta^t_{kv} &\sim \text{Normal}(0, 1),&
x^t_{s} &\sim \text{Normal}(0, 1).
\end{align*}
Restricting the mean values to zero implies that an average speaker in
the latent dimension is viewed as a neutral speaker and the average
intensities across the latent dimension correspond to the neutral
topic intensities. Setting the variance to one improves
identifiability of the model, as this fixes the possible within-topic
variability due to changes in the polarity of the speaker.

The parameters and latent variables are estimated for each session
separately by approximating their posterior distribution given the
priors and the observed data. The posterior of interest
$p(\bm{\theta}^t, \bm{\beta}^t, \bm{\eta}^t, \bm{x}^t|\bm{c}^t,
\bm{s}^t)$ is not available in closed-form and it is also
computationally intractable to directly obtain estimates from the
posterior. One thus usually resorts to approximation using variational
inference \citep{Blei+Kucukelbir+McAuliffe:2017}. Using variational
inference, the parameters and latent variables are estimated based on
an approximation of the posterior obtained by minimizing the
Kullback-Leibler divergence between the posterior and a variational
distribution. Point estimates are obtained by considering the
posterior means implied for the parameters and latent variables by the
fitted variational distributions.

This model specification and estimation scheme enables the derivation
of session-specific parameter and latent variable estimates which
captures differences and adaptations over time.  However, we are also
interested in linking topics as well as the latent space over time.
This is particularly relevant due to the identifiability issues
present for the TBIP model, i.e., neither the topics nor the latent
space are identifiable. To obtain a time-varying version of the TBIP
model, where the parameters are congruent across time, we make use of
a step-wise initialization approach across sessions where the
estimates obtained in the previous session are used as initial values
in the next session as far as possible when using a general purpose
optimizer to minimize the Kullback-Leibler divergence which requires
some initial values to be provided anyway.

In the first session, no previous results are available and
non-negative matrix factorization \citep[NMF;][]{Lee+Seung:1999} is
used to initialize the variational parameters for $\bm{\theta}^{1}$
and $\bm{\beta}^{1}$.  The input to the NMF algorithm is the
document-term matrix of the first session (i.e., $\mathbf{c}^{1}$),
which is approximated as the product of the topical prevalence matrix
$\bm{\theta}^1$ and topical content matrix $\bm{\beta}^1$, i.e.,
$\mathbf{c}^{1}\simeq \bm{\theta}^{1} \bm{\beta}^{1}$.  These two
matrices obtained with NMF are first ensured to only contain positive
entries and then the logarithm is taken. The resulting values are used
to initialize the means of the log-normal distributions used as
variational distributions for these parameters.

For each subsequent session with time index $t\ge 2$, the estimated
posterior means of the term prevalence vectors of the topics
$\beta^{t-1}_{k}$ are used to initialize the mean parameters of the
corresponding variational parameters in the TV-TBIP model. These
estimated prevalences are also used as input in the NMF algorithm
together with the document-term matrix $\bm{c}^t$ such that only the
initial values for the topic distributions $\bm{\theta}^{t}$ of the
speeches are determined in this step, i.e.,
$\mathbf{c}^{t}\simeq \bm{\theta}^{t} \bm{\beta}^{t-1}$ is used for
initialization.  In addition, the estimated values of
$\bm{\eta}^{t-1}_{k}$ are put forward as the means of the Gaussian
variational family of $\bm{\eta}^{t}_{k}$. Specifying the initial
values in this way ensures that the term prevalences of topics and the
topic-specific polarity scores are aligned across sessions preventing,
for example, label switching of topics and reversing of the latent
space\footnote{This modeling approach has also been
    pursued to study central bank communication in the euro area
    \citep[see][]{feldkircher2024}.}.  For more details on the model
estimation, we refer to the Online Appendix.

\section{40 Years of Political Discourse in the U.S.  Senate}\label{sec:results}
We use the Stanford University Social Science Data Collection database
for our analysis \citep{Gentzkow+Shapiro+Taddy:2018}. This database
was also used in GST and provides pre-processed text data on the
speech level from the United States Congressional Record, covering
Congress sessions 97 to 114 (1981--2017).  We use all speeches given
in the U.S.\ Senate and implement pre-processing steps similar to GST
and \citet{Vafa_etal_2020} to obtain session-specific document-term
matrices, using bigrams as tokens and retaining only those used by at
least 10 speakers per session. We also only retain speakers with at
least 24 speeches per session. Our model is fitted using 614,613
speeches given by 355 unique speakers, assuming 25 topics per
session. This is in line with GST who manually
identified 22 substantive topics based on their knowledge of the
database.  More details on data pre-processing and implementation of
the estimation are given in the Online Appendix.

\subsection{Average Partisanship}

\begin{figure}[t!]
  \centering
\begin{minipage}{.49\textwidth}
  \centering
   \includegraphics[width=\textwidth, trim = 0 0 0 0, clip]{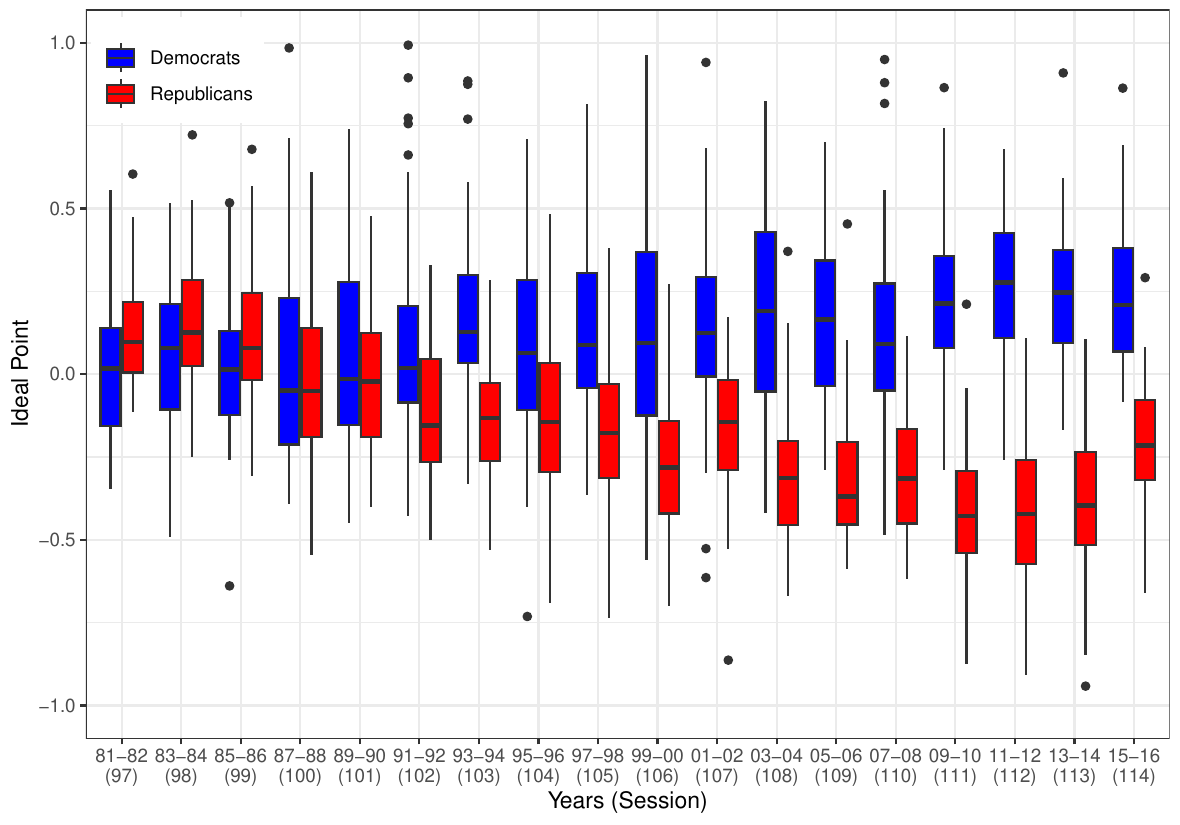}
\end{minipage}
\begin{minipage}{.49\textwidth}
   \centering
   \includegraphics[width=\textwidth, trim = 0 0 0 0, clip]{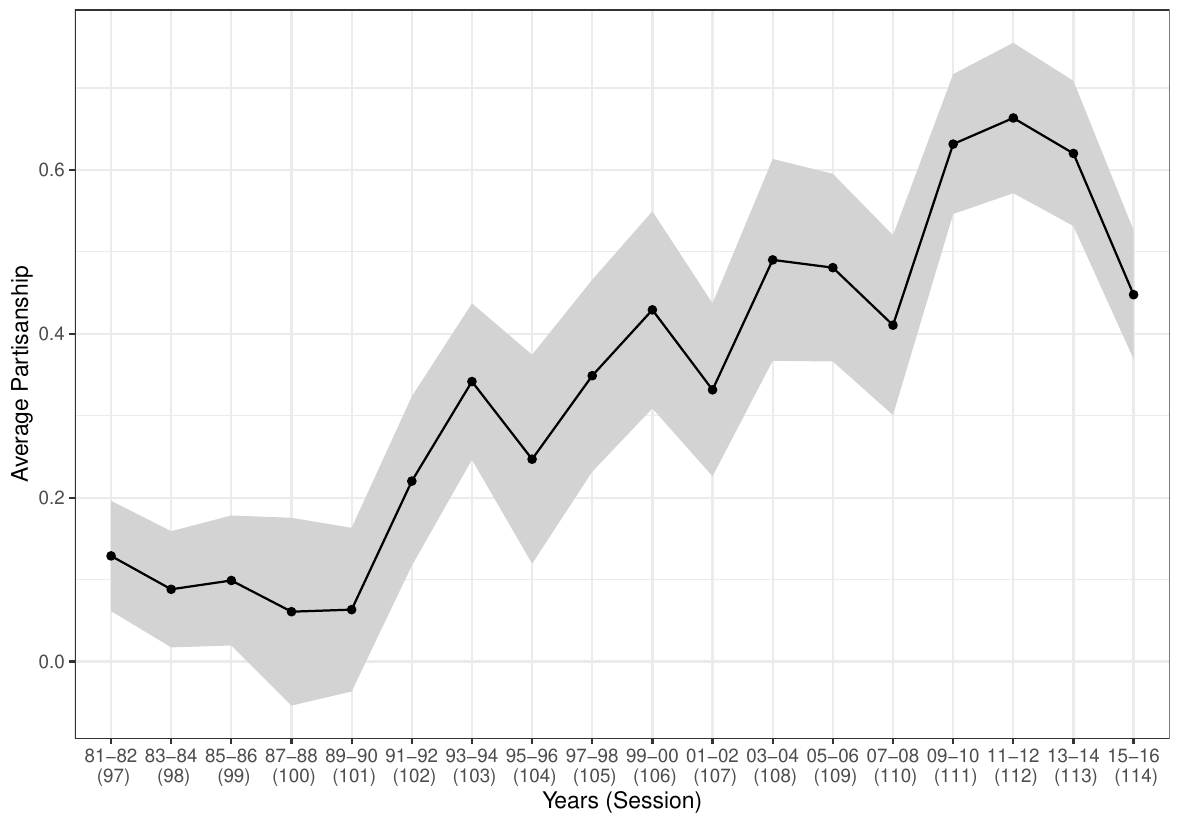}
\end{minipage}
\caption{Party- and session-specific ideal point distributions
  represented as box-plots (left). Estimated average partisanship over
  the years together with approximate pointwise 95\% confidence
  intervals (right). 
  \label{fig:IP}}
\end{figure}

We inspect the session- and party-specific distributions of the
estimated ideal points to assess if the latent scale in fact
differentiates between Democrat and Republican speakers for each
session.  Figure~\ref{fig:IP} (left) summarizes these ideal point
distributions based on box-plots. Note that the ideal points are
a-priori assumed to be standard normally distributed with the mean
zero indicating a neutral speaker. The values of the estimated ideal
points may thus be interpreted as reflecting the degree of
partisanship of the speakers with the distance of the ideal point from
zero indicating the deviation of the speaker from a neutral speaker in
the unit of standard deviations.  For each session, a pair of
box-plots is displayed for the two parties. This provides a means to
compare the party-specific locations based on the medians and assess
the overlap based on the boxes.  Clearly the boxes overlap strongly in
the first sessions, while boxes do not overlap after session 105.

To study the evolvement of average partisanship across Congress
sessions, we determine an estimate of the average partisanship between
the parties over time by aggregating the point estimates of the
ideological positions $\hat{x}^t_{s}$ of the speakers across parties.
For each session, we calculate the mean of the estimated ideal points
separately for the members of each party and take the
difference. I.e., the average session-specific partisanship for
session $t$ is determined using
\begin{equation}
  \bar{\pi}^t = \left|\frac{1}{N^t_{R}}
                \sum_{s \in I^t_{R}} \hat{x}^t_{s} -
                \frac{1}{N^t_{D}}
                \sum_{s \in I^t_{D}} \hat{x}^t_{s}\right|,
\end{equation}
where $N^t_{R}$ and $N^t_{D}$ are the number of Republican and
Democrat speakers with ideal point estimates available for session $t$
and $I^t_{R}$ and $I^t_{D}$ denote the index sets for those speaker
groups. This approach neglects the uncertainty of the
  Bayesian point estimates and treats the estimates of these latent
  variables as if they were observed. Alternatively, a fully Bayesian
  approach could also be pursued.

Figure~\ref{fig:IP} (right) depicts the evolvement of these average
partisanship estimates over time and corroborates the key results of
GST.  The gray shaded area represents the approximate pointwise $95\%$
confidence intervals. These confidence intervals are obtained using
$\bar{\pi}^t \pm 1.96 \times
\sqrt{\frac{(\hat{\sigma}^t_D)^2}{n^t_D}+\frac{(\hat{\sigma}^t_R)^2}{n^t_R}}$,
where $\hat{\sigma}^t_D$ and $\hat{\sigma}^t_R$ are the sample
standard deviations of the estimated point estimates for the ideal
points for each party. In line with GST, the TV-TBIP model does not
detect any noteworthy partisanship in the speeches given in the U.S.\
Senate before the 1990s. However, at the beginning of the 1990s, the
estimated average partisanship between Democrats and Republicans
increases rapidly and peaks around 2010. This pattern of evolvement of
partisanship closely follow the results obtained using the supervised
model proposed in GST \citep[see][Figure~2 on
p.~1321]{Gentzkow_etal_2019}. The Online Appendix
  provides results of a simulation study where the same estimation
  method is used for analysis.  New document-term matrices for each
  session are generated based on the estimated posterior mean
  estimates and varying the distribution of ideal points across
  parties and sessions to induce no, fixed and increasing party
  differences as well as re-using the estimated ideal points for the
  speakers across sessions. These results confirm the suitability of
  the proposed approach to infer the evolvement of partisanship across
  sessions.

In line with GST, we also validate our aggregate measure of
partisanship by comparing it with the first dimension of the
DW-Nominate scores estimated from voting data \citep[see,
e.g.,][]{Poole:1985}. In particular, we take the difference between
the average DW-Nominate scores of Republicans and Democrats. We find a
correlation of $0.77$ indicating that our text-based average
partisanship measure captures essentially the same effect over time as
the vote-based DW-Nominate scores. The Online
Appendix provides more details on the timely evolvement of ideological
positions on speaker level.

Our model tracks the timely evolvement of ideological
positions on individual speaker level. Due to the pronounced
partisanship from session~105 onwards, we focus in the following
analysis on results from session~105 and later (see also the Online
Appendix for more details). We determine the average ideal point score
for each speaker across all sessions they were in the Senate during
this time period. These results indicate that the most liberal
Democrats in this time period are Byron Dorgan, Dale Bumpers, Thomas
Harkin, Christopher Murphy and Paul Wellstone. The same analysis for
the Republican party shows that TV-TBIP identifies Jesse Helms, a
Senator from North Carolina, as the most conservative Republican. This
is in line with how this Senator is depicted in the media, e.g., the
New York Times \cite[see][]{Helms1} stated that Helms was ``bitterly
opposed'' to federal financing for research and treatment of AIDS
which he believed was God's punishment for homosexuals \citep[see,
e.g.,][]{Helms}. According to the DW-Nominate score, Jesse Helms is
also ranked as the most conservative Republican for, e.g.,
sessions~106 and 107.

\subsection{Temporal Change, Drivers of Partisanship and Partisan Phrases}
\label{sec:topics}
GST requires manual specification of the topics based on domain
knowledge.
TV-TBIP facilitates the inference of topics in a data-driven manner,
thereby enabling an extension where topical content can change over
time.
\subsubsection{Temporal Change of Topical Content}
Topics are usually characterized by their top most frequent terms, and
in the following, we refer to the topic-specific term distribution
which is used by a neutral speaker, i.e., a speaker with ideal point
$x^t_{s}=0$, as the neutral topic.  Table~\ref{tab:topics} displays
the five most frequent terms of the neutral topics for the first and
the last session considered. The sessions at both ends of the time
frame are shown to highlight the overall evolvement of each of the
topics over time. 

\input{tableneutral}

\begin{figure}[t!]
  \centering
  \includegraphics[width=0.32\textwidth]{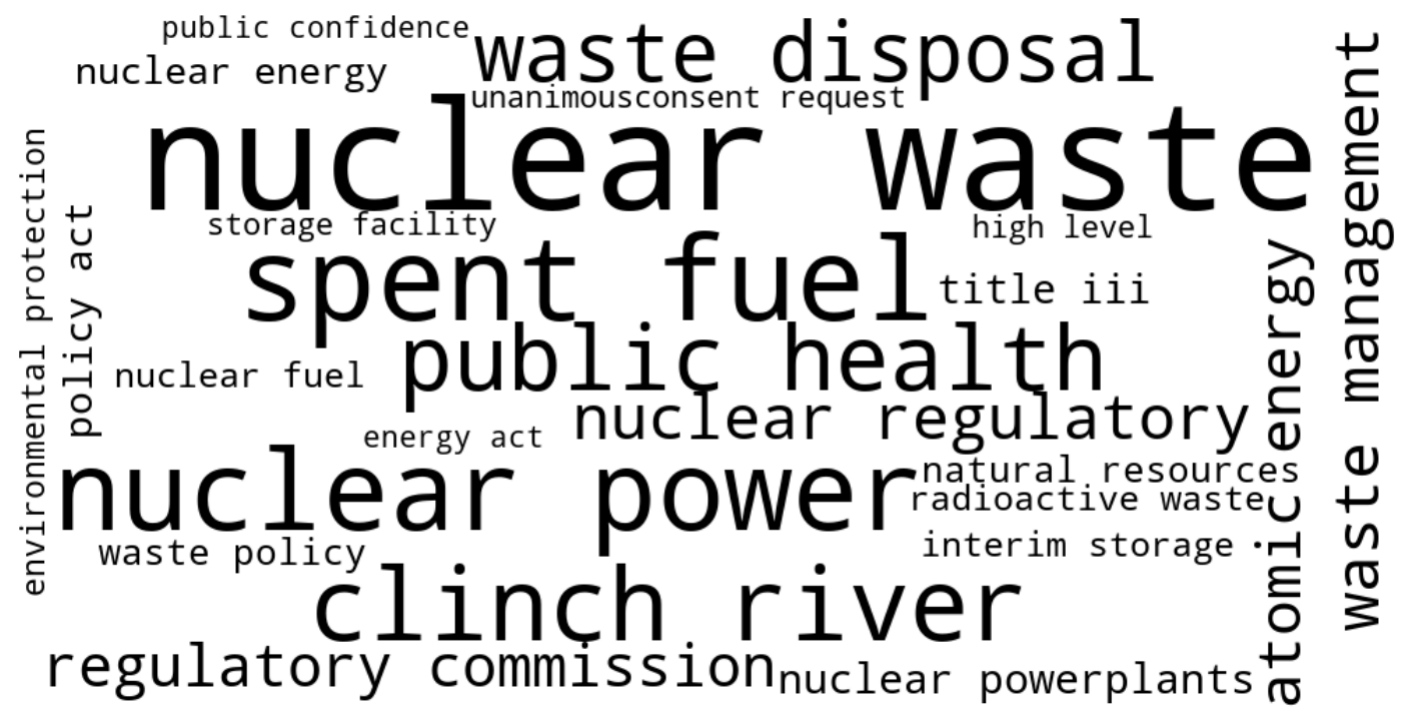}
  \includegraphics[width=0.32\textwidth]{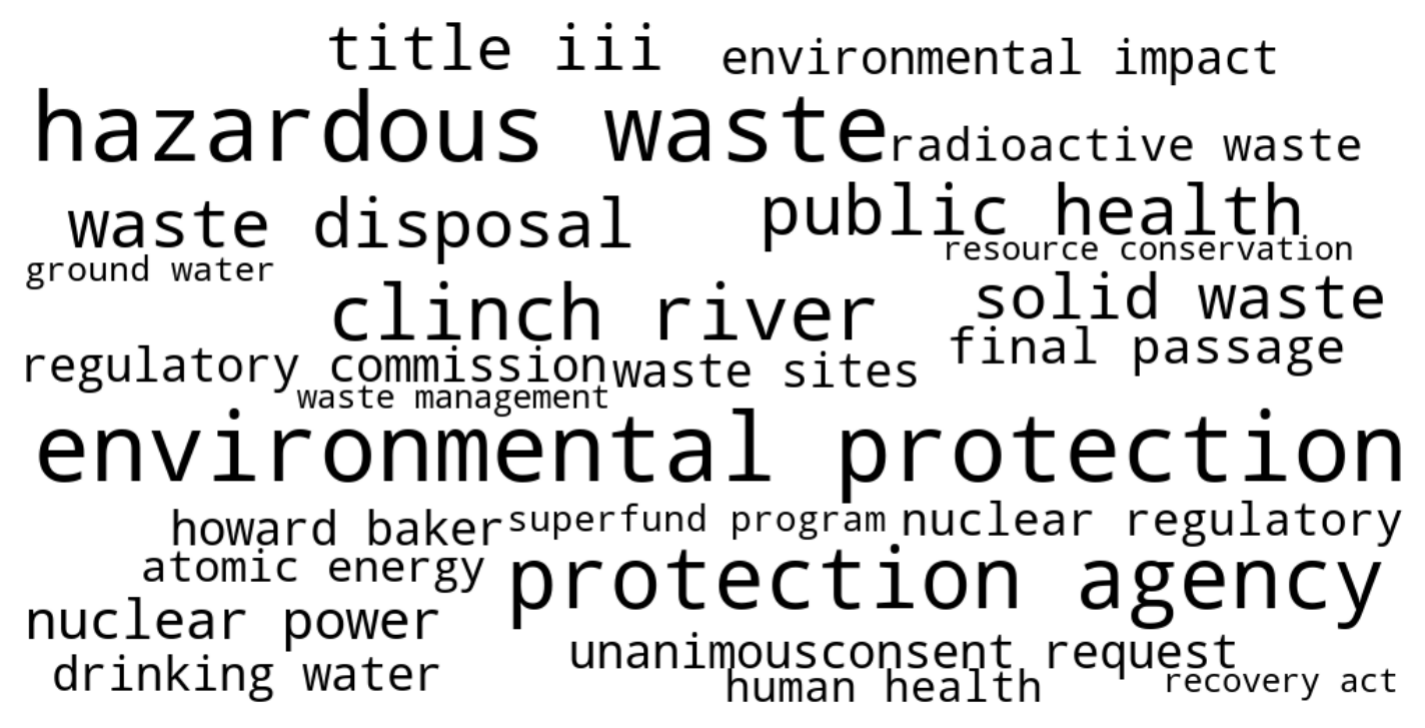}
  \includegraphics[width=0.32\textwidth]{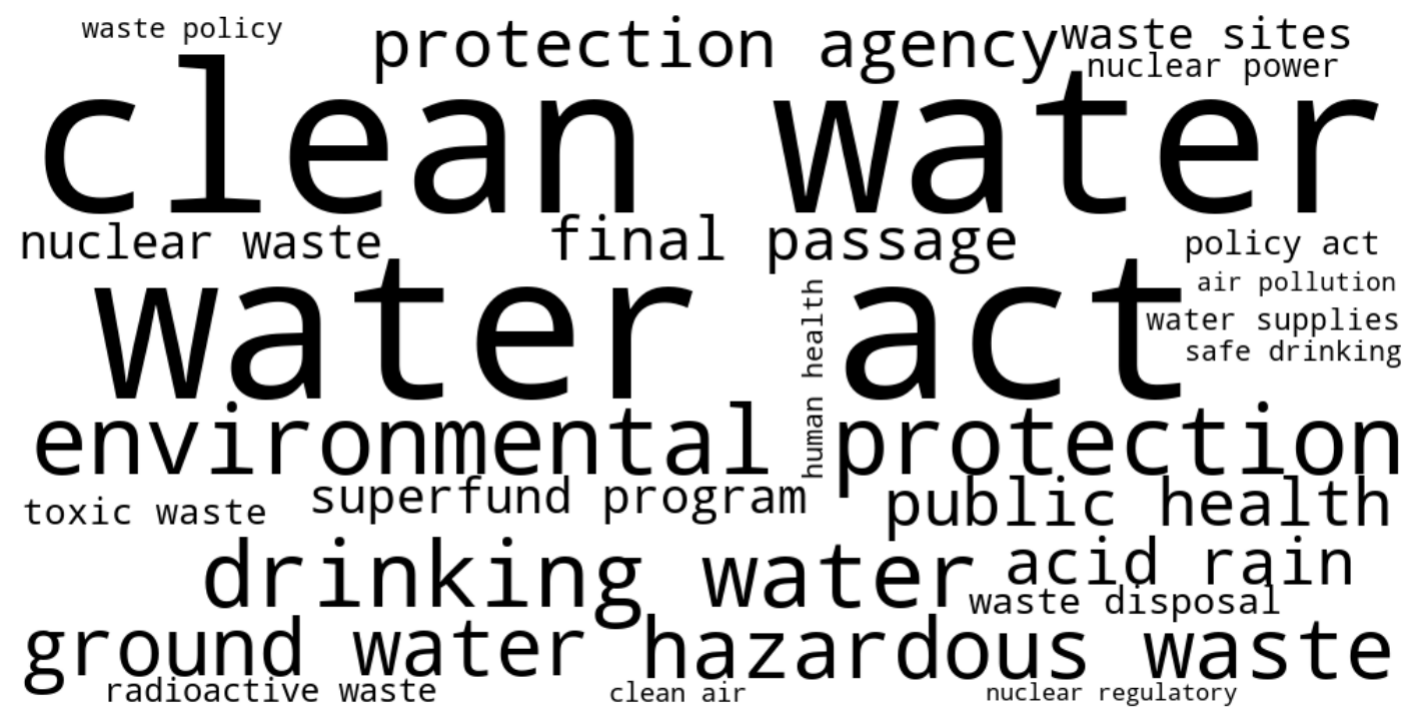}
  
  \includegraphics[width=0.32\textwidth]{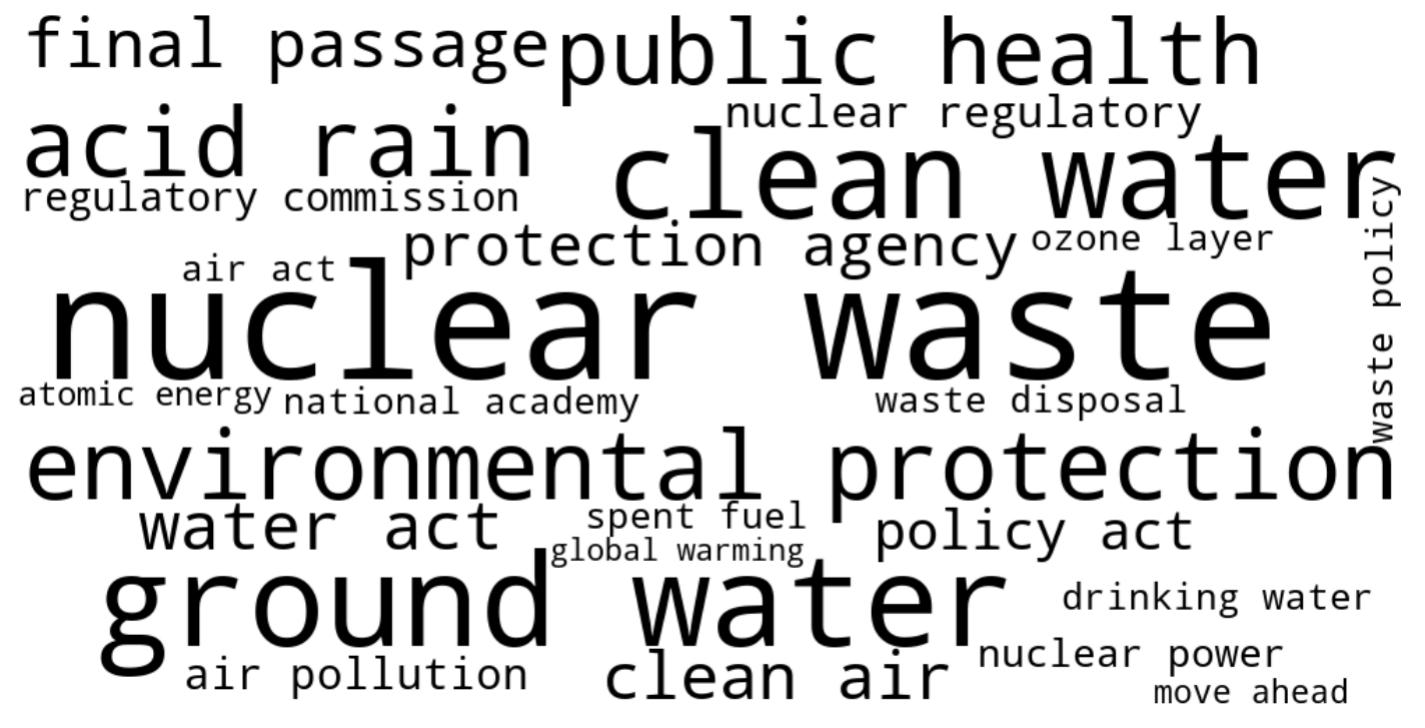}
  \includegraphics[width=0.32\textwidth]{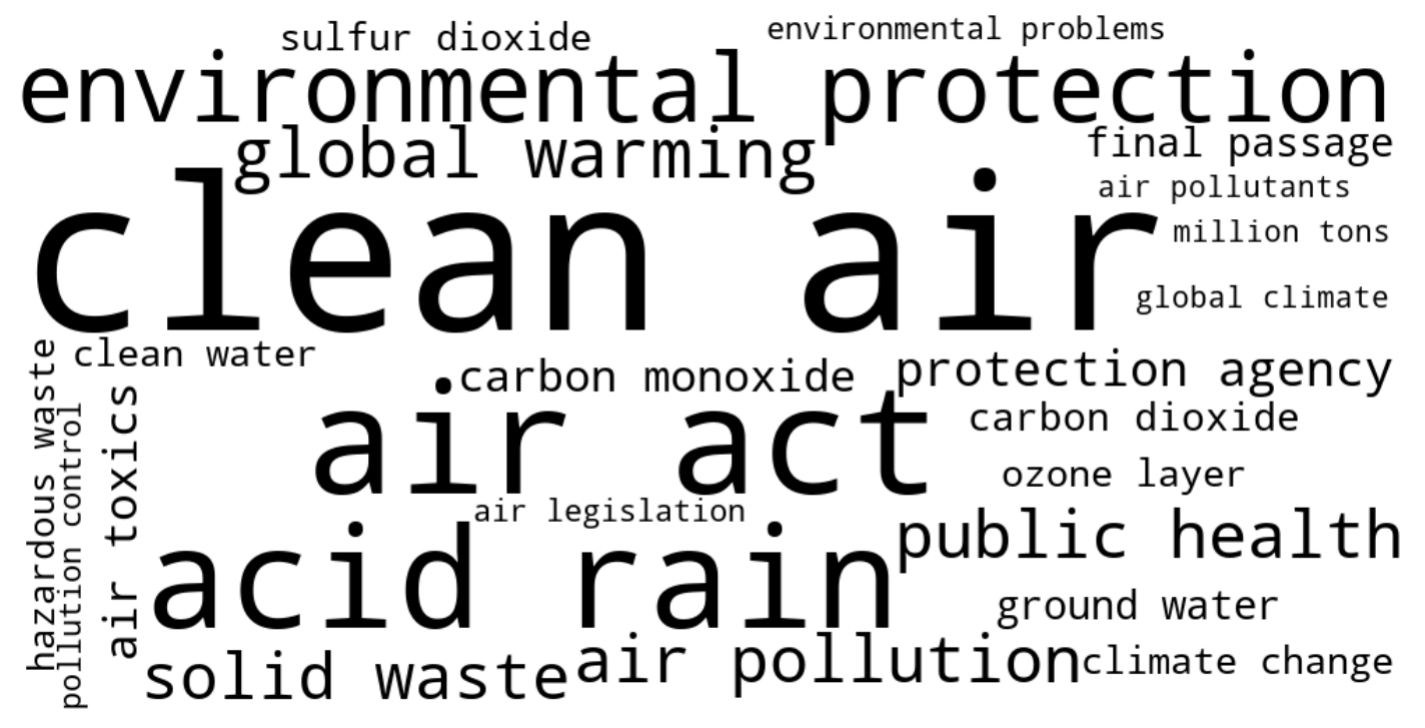}
  \includegraphics[width=0.32\textwidth]{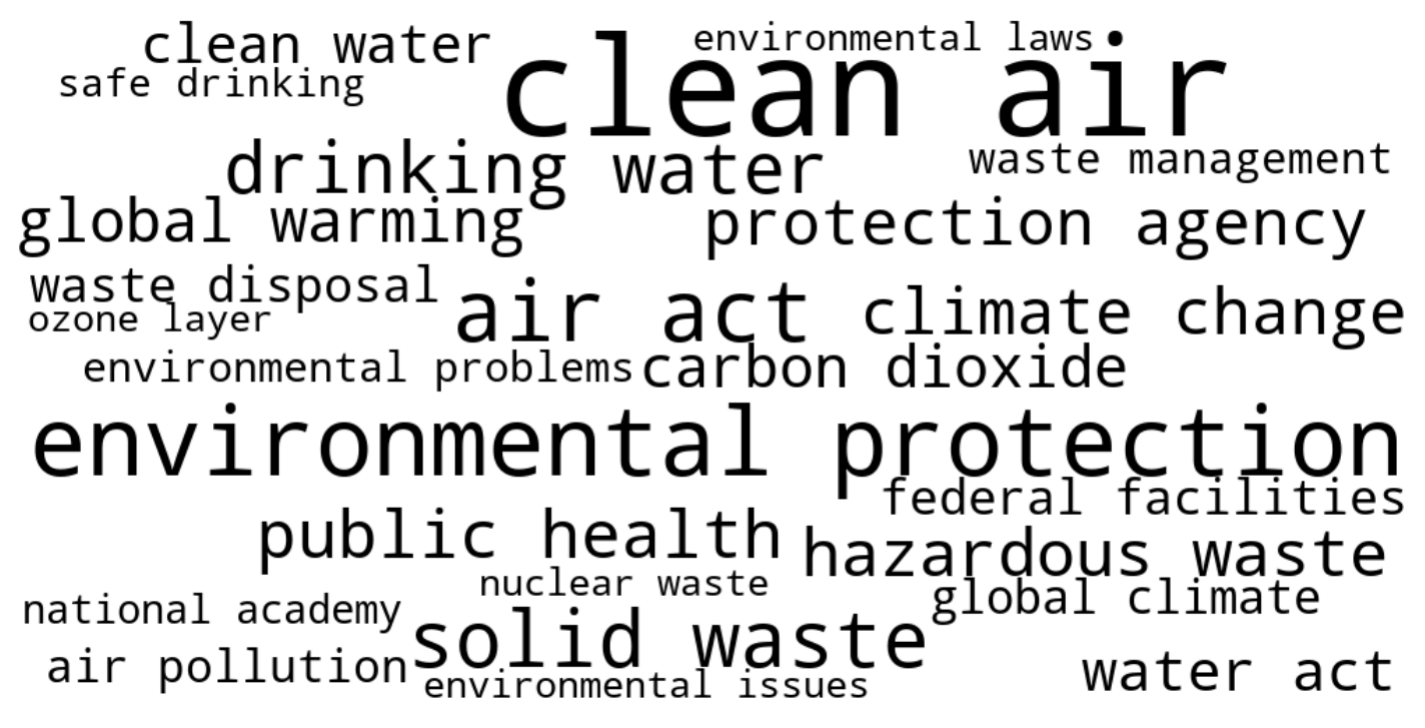}
  
  \includegraphics[width=0.32\textwidth]{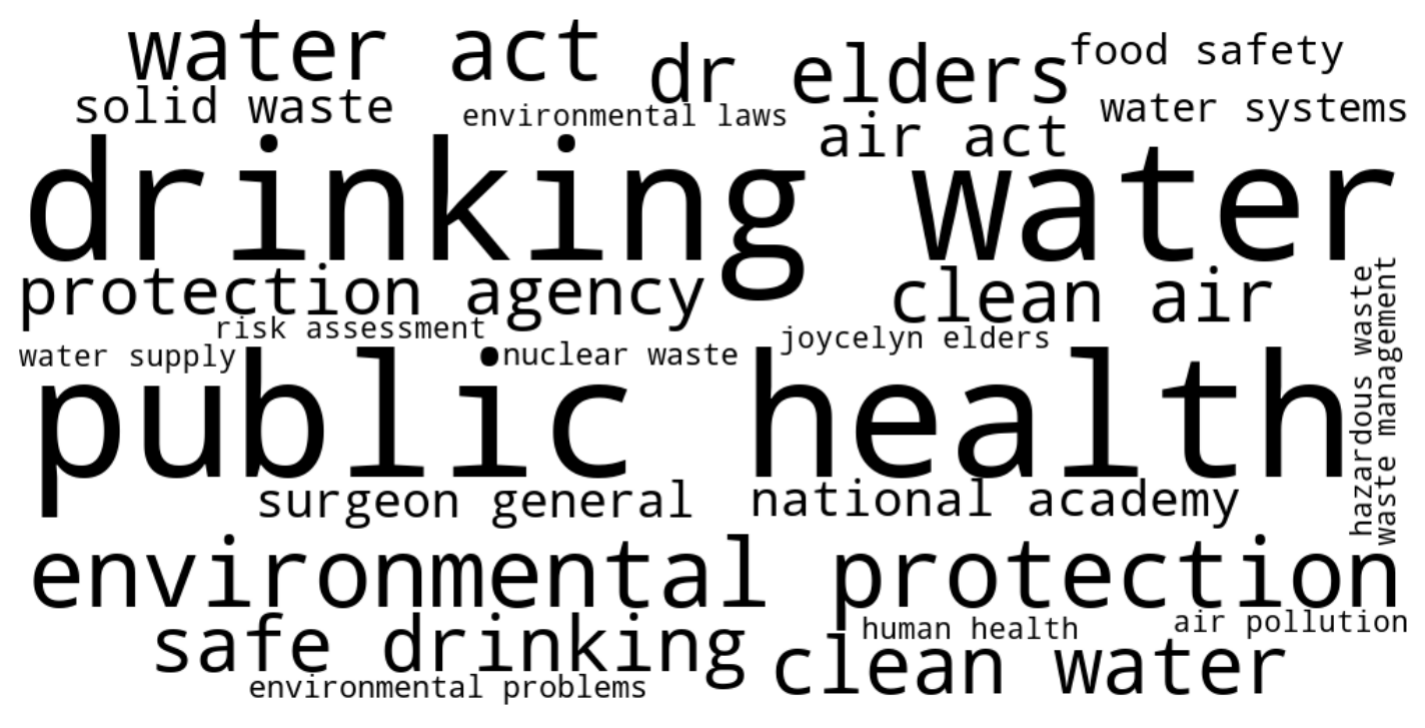}
  \includegraphics[width=0.32\textwidth]{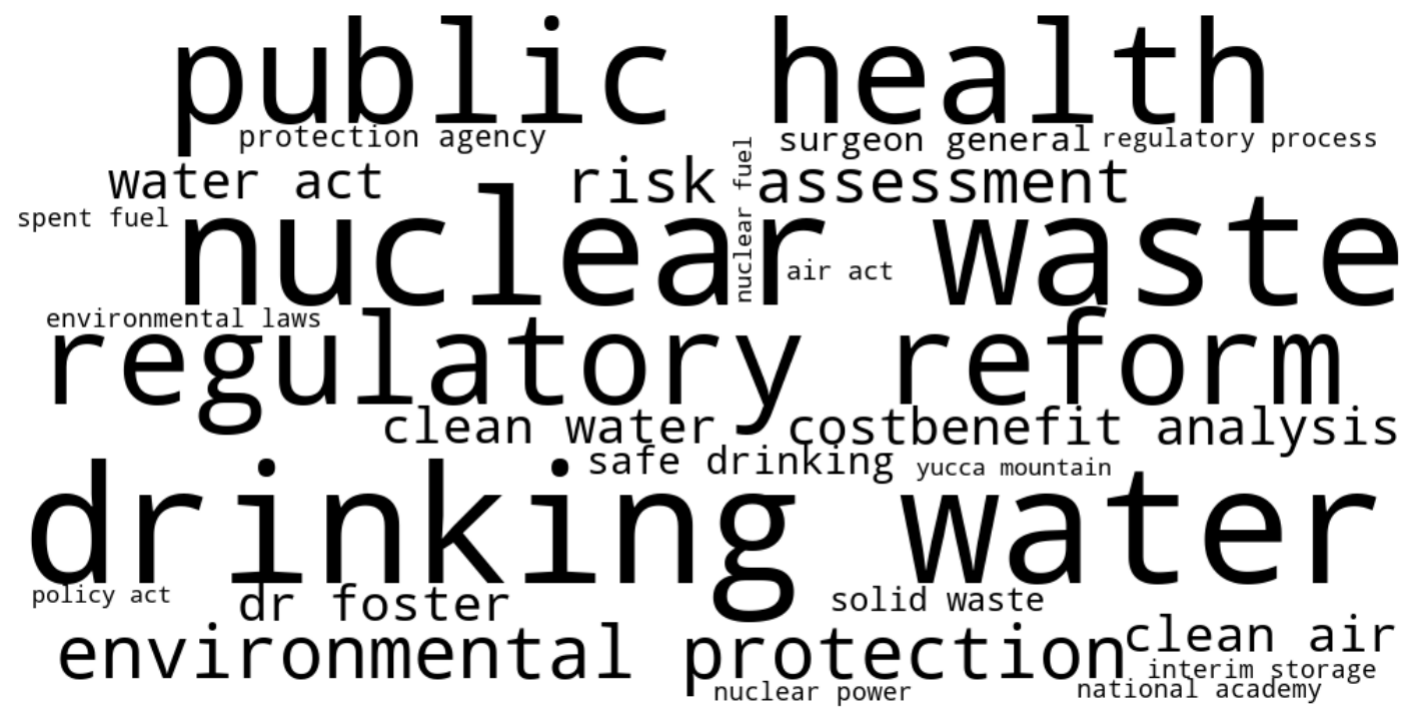}
  \includegraphics[width=0.32\textwidth]{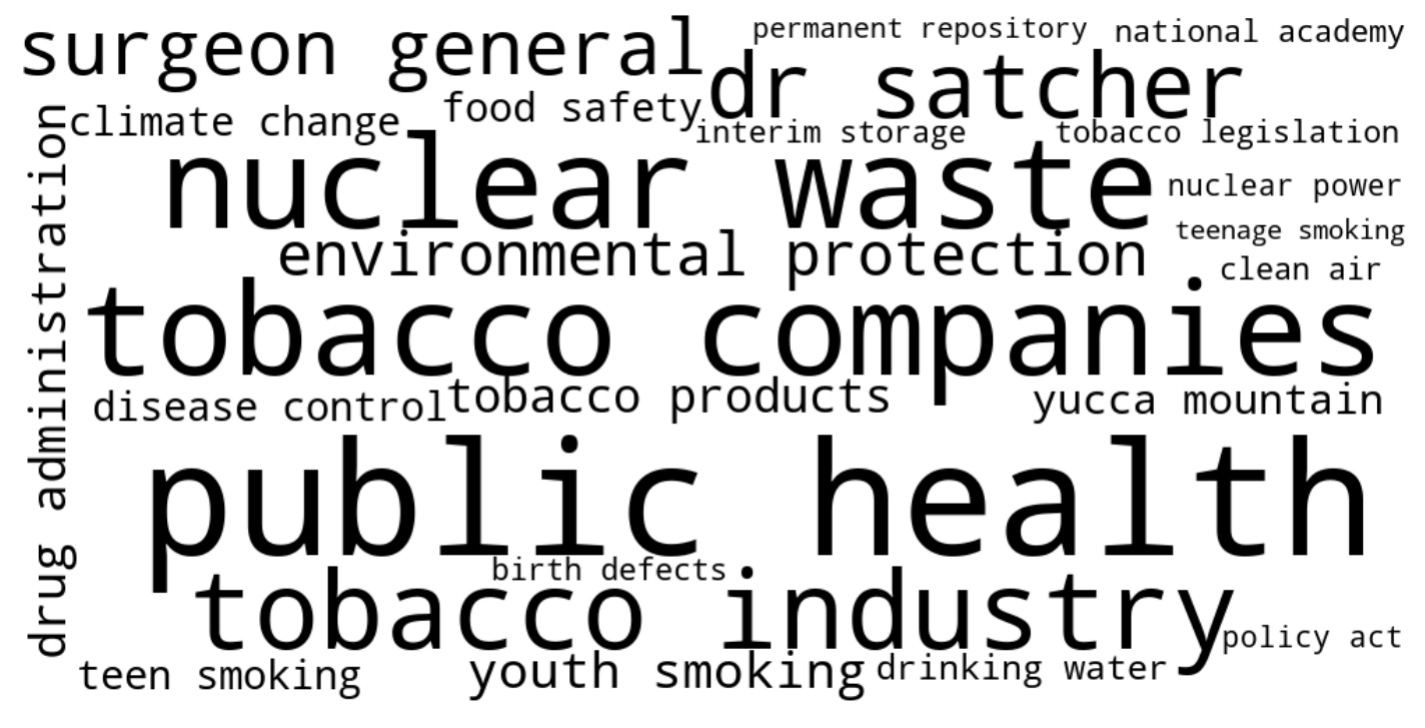}
  
  \includegraphics[width=0.32\textwidth]{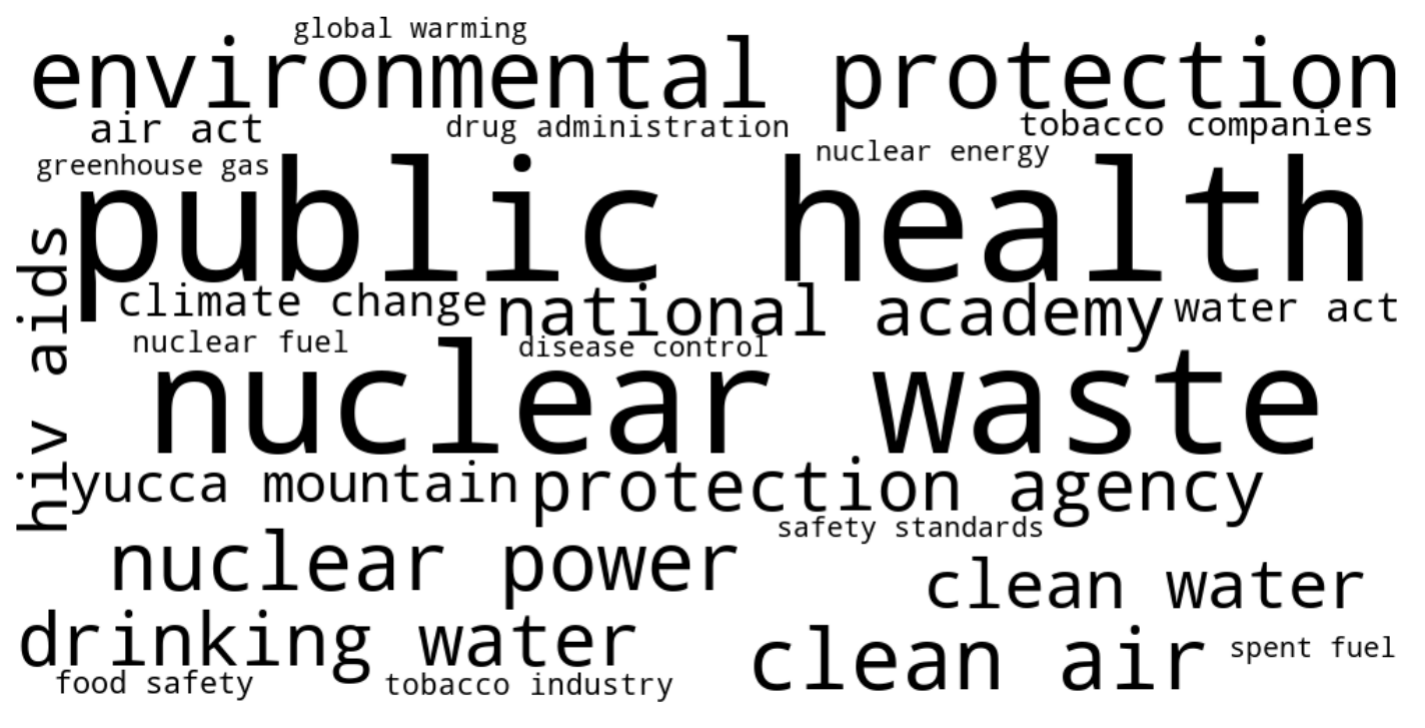}
  \includegraphics[width=0.32\textwidth]{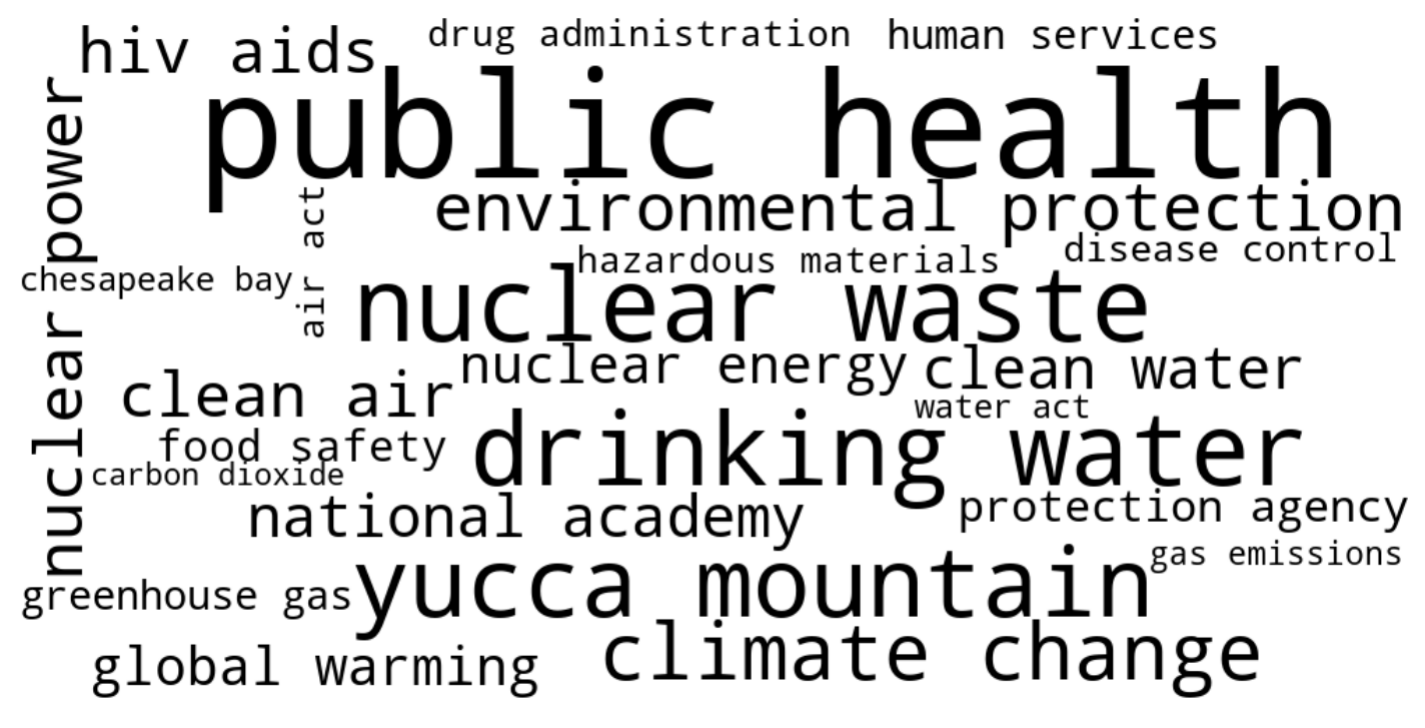}
  \includegraphics[width=0.32\textwidth]{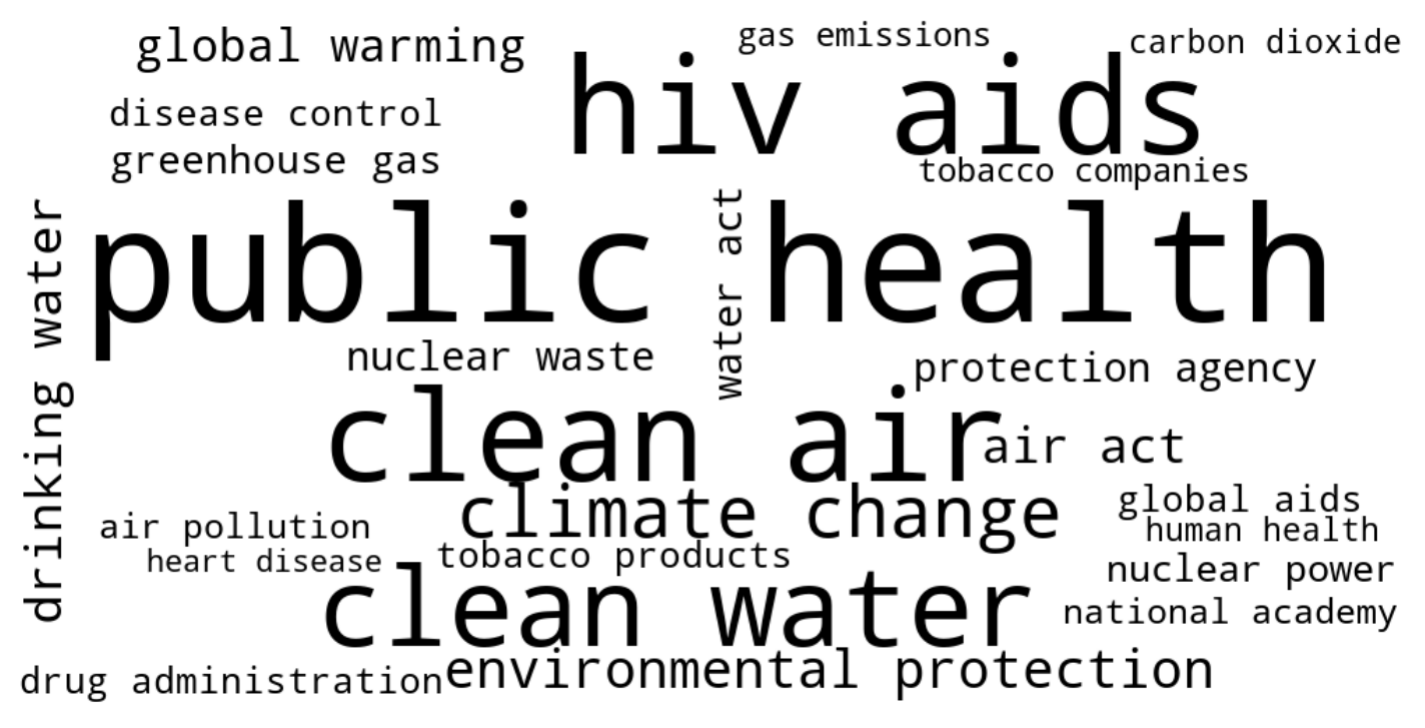}
  
  \includegraphics[width=0.32\textwidth]{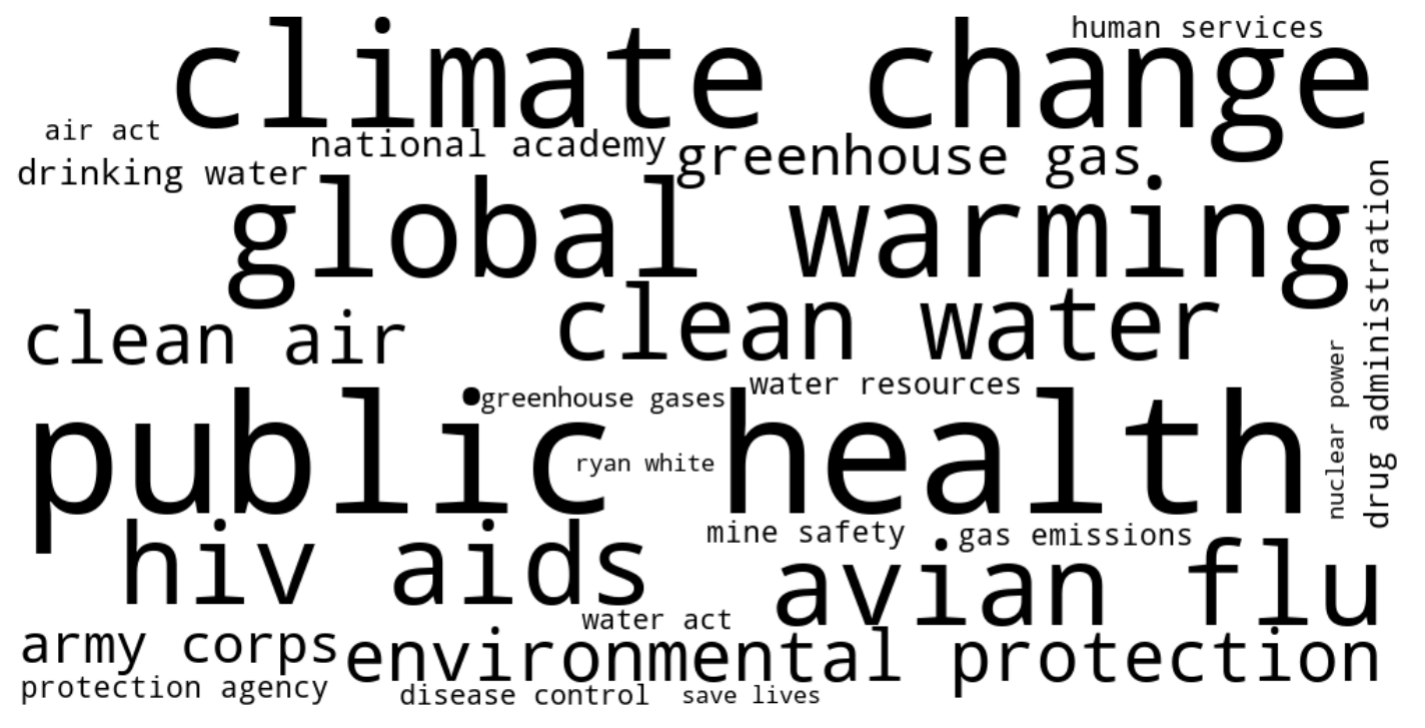}
  \includegraphics[width=0.32\textwidth]{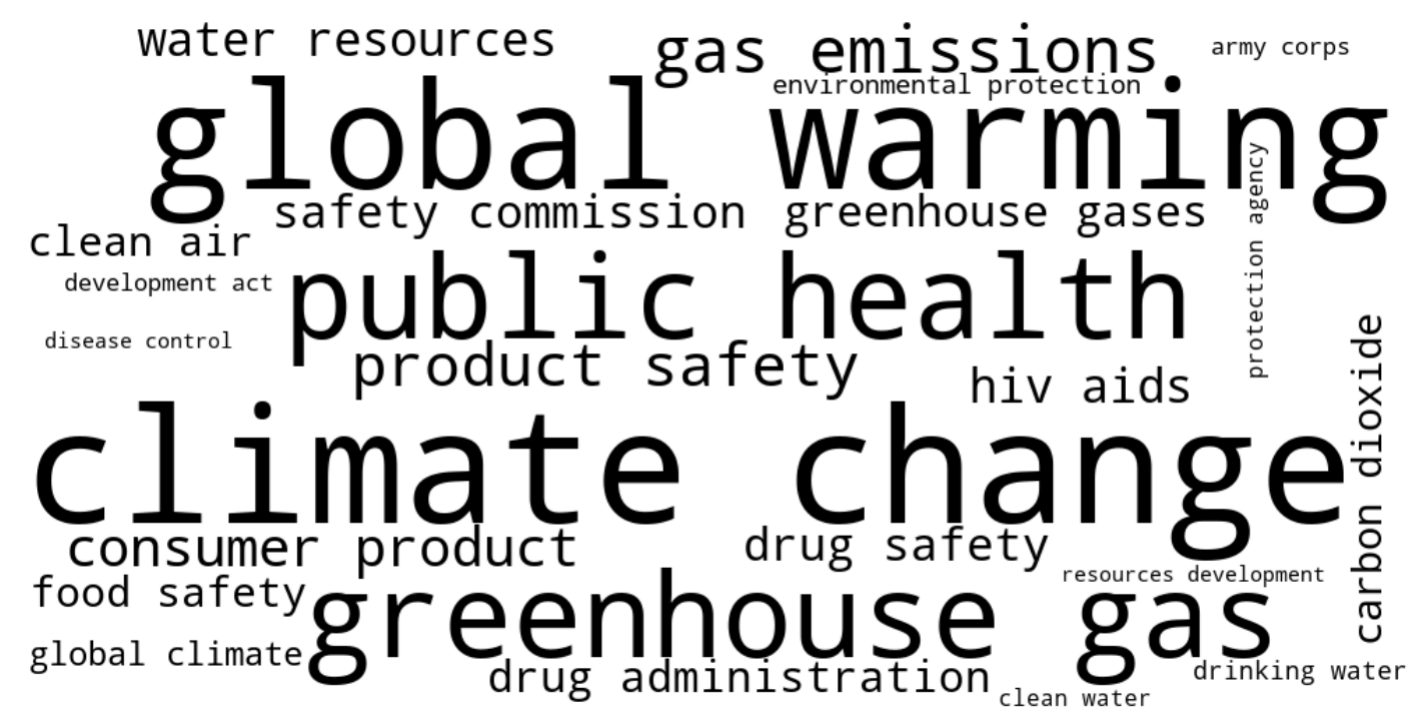}
  \includegraphics[width=0.32\textwidth]{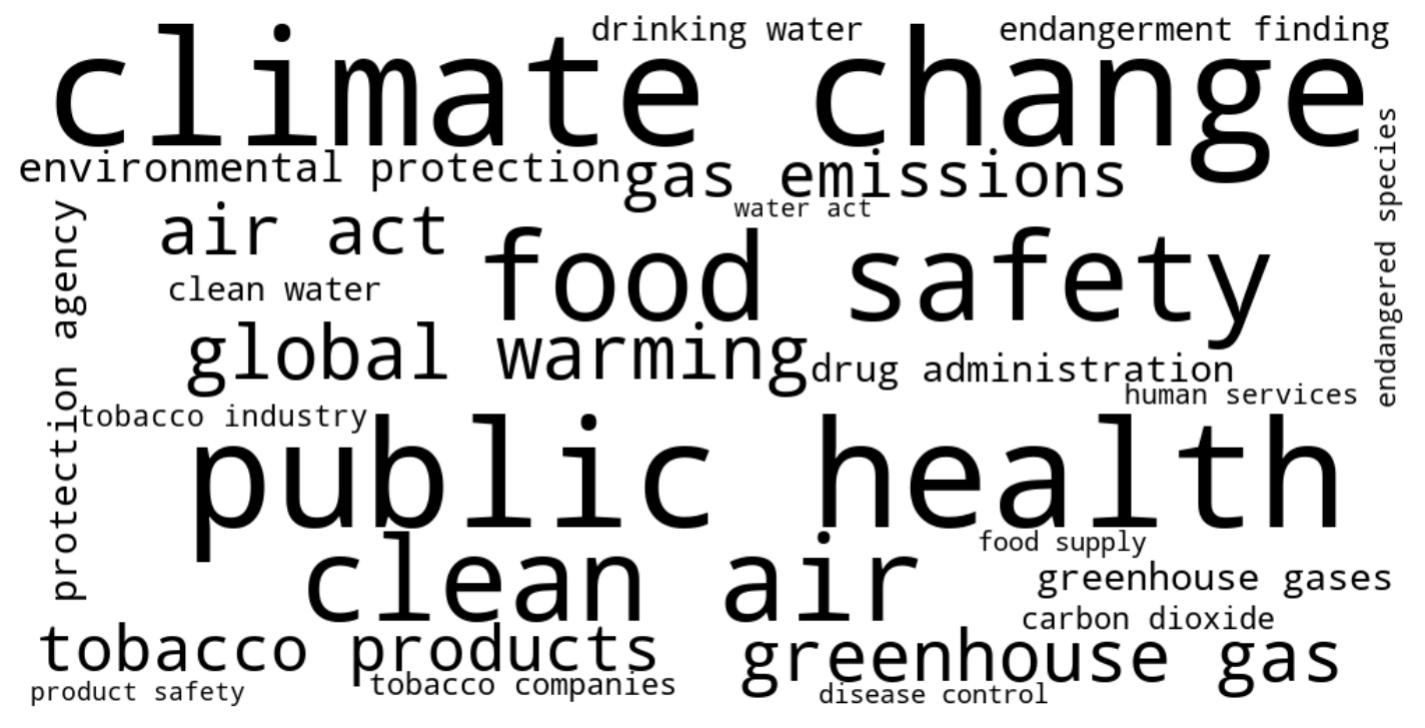}
  
  \includegraphics[width=0.32\textwidth]{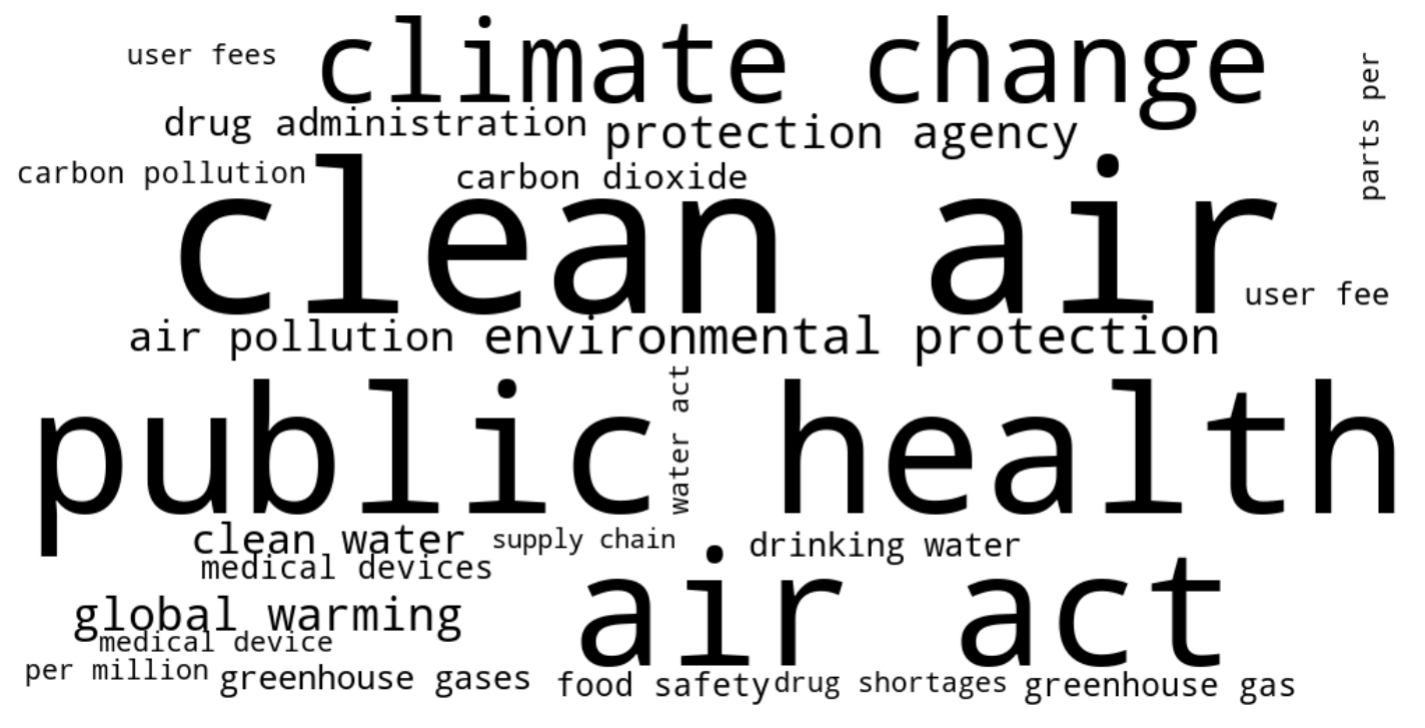}
  \includegraphics[width=0.32\textwidth]{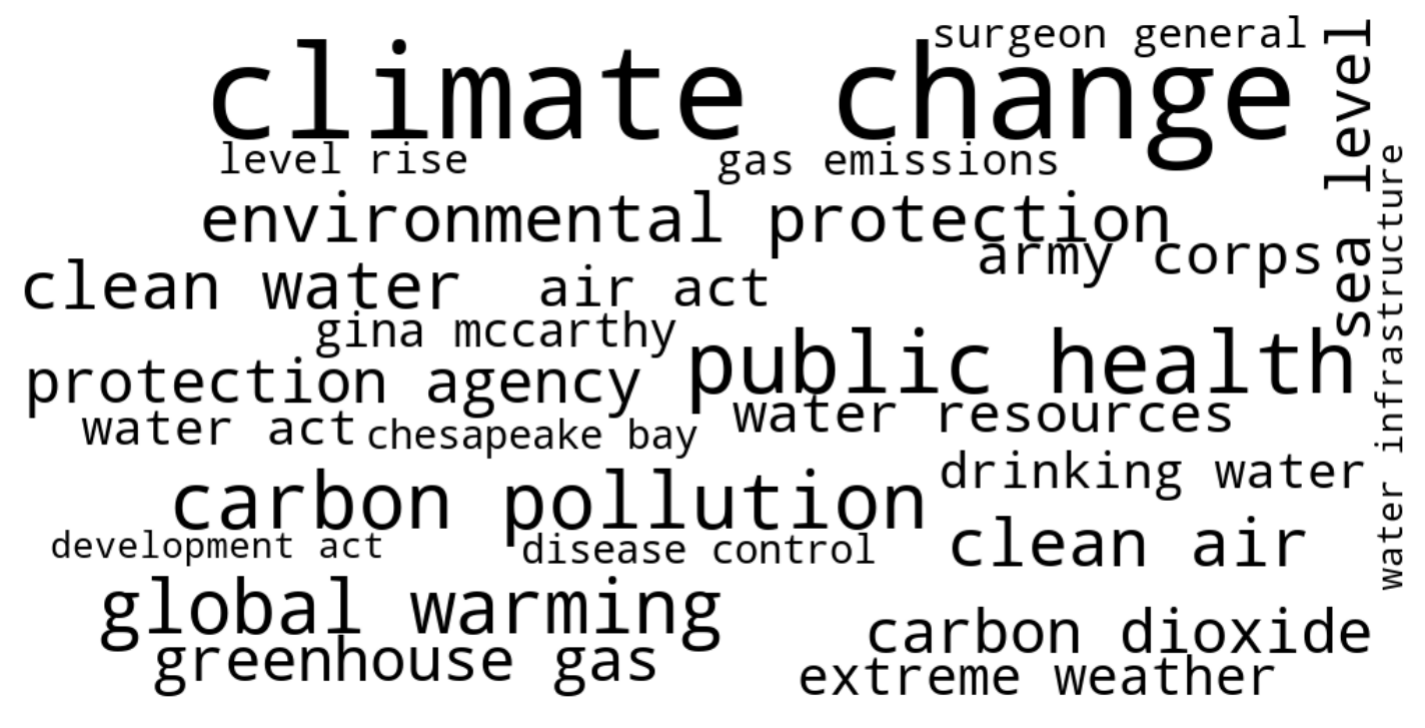}
  \includegraphics[width=0.32\textwidth]{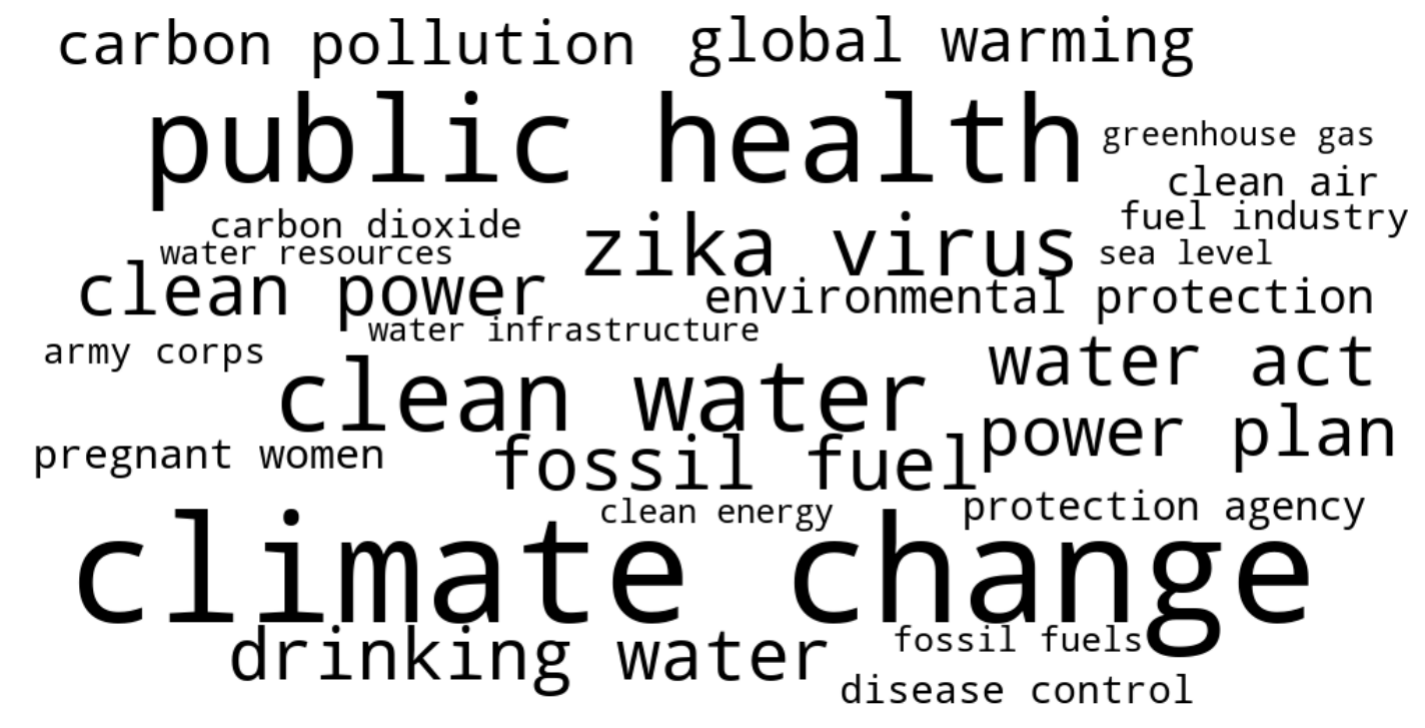}
  \caption{Term appearance rates of Topic~11, the climate
    change/public health topic, across sessions~97 to 114, starting
    with session~97 on the top left and proceeding
    row-wise. \label{fig:topic11}}
\end{figure}

Table~\ref{tab:topics} indicates that the first topic is mainly
concerned with the \emph{United States}. This bigram has by far the
highest appearance rate (approximately $0.3$ for both
sessions\footnote{Indeed, we observe a similarly high value for all
  sessions.}) across all bigrams in the vocabulary. 
The additional bigrams listed as having the highest appearance rates
for Topic~1 indicate that this topic is about the United States and
their concerns with other states such as \emph{international trade,
  foreign relations, trade agreements}. Selectively characterizing
some other topics, one can discern that Topic~8 is about foreign
policies in the Middle East and its most frequent terms changed from
\emph{saudi arabia, foreign policy} to \emph{al quaeda, islamic state}
from the first to the last session; Topic~12 is about taxes in
general, but moves from a discussion on tax cuts to focus more on
taxation and the middle class from the first to the last session.  Of
specific interest is also Topic~11, which represents a climate
change/public health topic. For this topic, nuclear waste is a
prevalent term in session~97, which changes to climate change being a
prevalent term in session~114.  Figure~\ref{fig:topic11} displays the
evolvement of that topic using word clouds across all sessions. It
changed from the discussion of nuclear waste over acid rain to climate
change and public health related issues.

\begin{figure}[t!]
\centering
\includegraphics[width=0.48\textwidth, trim = 5 20 5 20, clip]{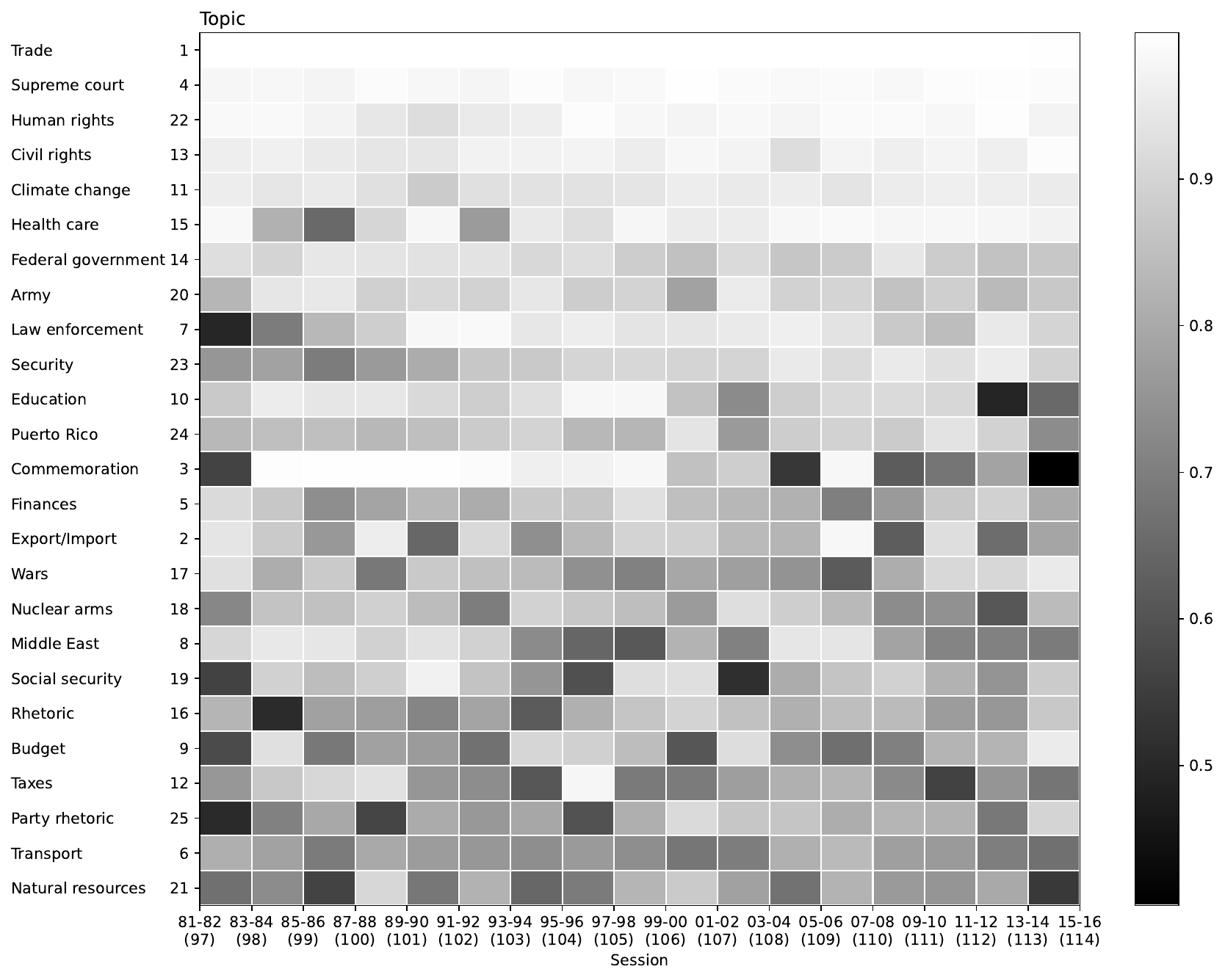}\hspace*{0.02\textwidth}
\includegraphics[width=0.48\textwidth, trim = 5 20 5 20, clip]{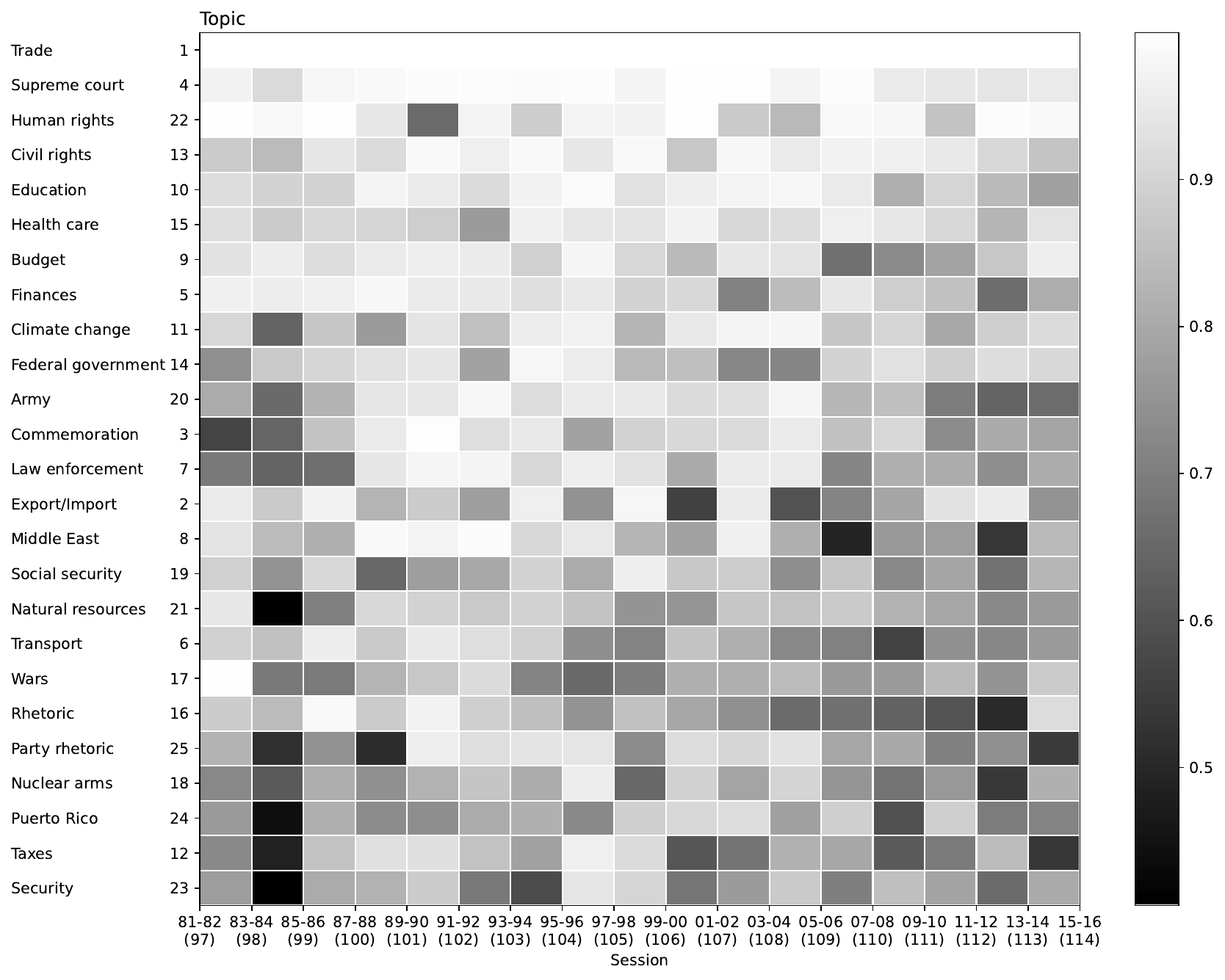}
\caption{Left: Similarity of the topics for consecutive sessions based
  on the term compositions of the neutral topics. Darker values
  indicate a higher change in term composition. Right: Similarities
  between the term compositions of the positive and negative
  ideological topics for each session. Darker values indicate a higher
  polarization for that topic in a specific
  session. The most frequent terms associated with the
    each of the topics can be found in Table~\ref{tab:topics}.
    \label{fig:topics-similarity}}
\end{figure}

To assess stability of topic composition in a systematic way, we determine the cosine
similarity between the estimated term intensities of the same topic
for two consecutive sessions. Figure~\ref{fig:topics-similarity} on
the left displays a heat map of these cosine similarities. The topics
are sorted by their mean cosine similarity with the most stable one
being on the top and the most volatile one at the bottom. For this
comparison, the term intensities of the neutral topics were used.
On the top, we find Topic~1 (the United States topic) which is the
most stable topic over time. Other topics with stable topical content
over time are Topic~14, which is concerned with the federal and local
governments, and Topic~19, which is about social security. Among the
three most volatile topics we find again Topic~11 (see
Figure~\ref{fig:topic11}).  In addition, the largest change in term
composition of topics occurs for Topic~23 from session~112 to 113 with
still quite a substantial change in term composition to the subsequent
session~114. Inspecting the most frequent terms for Topic~23 for
sessions~112 and 113 indicates that the term \emph{national security}
is substituted by \emph{homeland security} and even more important
\emph{immigration} issues started to be also raised in that
topic\footnote{Note that similar observations have also been discussed
  in
  \href{https://www.everycrsreport.com/reports/R43097.html}{https://www.everycrsreport.com}.}.

\subsubsection{Topics as Drivers of Partisanship}
In the following we analyze which topics are main drivers of partisanship.
We create again a heat map of the
cosine similarity matrix, this time comparing the term frequencies of
positive and negative topics for each topic and session. The term
prevalences of the positive and negative topics are determined using
an ideal point of $1$ (i.e., liberal) and $-1$ (conservative). The
results are displayed in Figure~\ref{fig:topics-similarity} on the
right. Topics are again sorted by their mean cosine similarity with
the least polarizing being on the top and the most polarizing one at
the bottom.

We find that Topic~1 is the most ``neutral'', i.e., the topic where
the differences between the terms used by liberal and conservative
speakers are smallest.  In addition we also observe low discordance
for Topic~14 which is about the federal government and Topic~17 which
is about war veterans.  Topic~11 on climate change/public health is
close to the bottom of the heat map in
Figure~\ref{fig:topics-similarity} on the right, indicating that there
is less congruence in the wording of Democrats and Republican when
talking about these issues.  The same holds for Topic~21, another
energy topic which is ranked as the third least congruent topic. Only
Topic~9 (which is about government shutdowns and budget control) and
Topic~16 (which appears to be a derivation thereof) have a lower
concordance.

\subsubsection{Partisan Phrases}
GST present estimated partisan phrases for selected periods, i.e.,
phrases which allow to differentiate most between a Republican and a
Democrat speaker (see Table~1 on p.~1325 which provides the 10 most
important partisan phrases for every 10th session).  In the TV-TBIP
model, partisan phrases emerge on the topic level because the term
distributions of the topics are influenced by the polarity scores and
the ideal points of the speakers. To identify partisan phrases on
topic level, we inspect the term compositions of the topics for a
liberal speaker with ideal point value $1$ and a conservative one with
ideal point value $-1$. More formally, we compare the term
distribution $\bf{\beta} \exp\{\bf{\eta}\}$ with
$\bf{\beta} \exp\{\bf{-\eta}\}$. Table~\ref{tab:tableposneg} displays
the most frequent terms of each topic for these speakers.

In this analysis, our focus is on session 114, which occurs at the end
of the observation period and where the latent dimension distinctly
enables discrimination between Republican and Democrat speakers.
In line with GST, who identify \emph{puerto rico} as a partisan
phrase, we also see that \emph{puerto rico} is among the most frequent
positive or negative terms for Topics 1, 17 and 24. E.g., for Topic~17
we estimate an appearance rate of $0.049$ for Democrats and $0.03$ for
Republicans. GST find rates of 42 (R) and 79 (D) per 100,000
phrases. Further, also in line with GST, \emph{religious freedom}
appears among the most frequent term only for Republicans. GST find
this term to appear 34 times per 100,000 phrases for Republicans and 4
times for Democrats.  The TV-TBIP results also highlight the topic
where this word appears, i.e., Topic~22.

\input{tableposneg}

In addition to GST, e.g., our results indicate that for Topic~21 --
which is about natural resources and energy -- a speaker with a
positive, i.e., a liberal, ideal point uses terms like \emph{energy
  efficiency} or \emph{clean energy} when talking about this topic
while a conservative one uses terms like \emph{keystone xl} (a
pipeline project by TC Energy) or \emph{energy security}.  For
Topic~5, which is about monetary policy, we find on the liberal side
terms like \emph{financial crisis} and \emph{consumer protection} but
on the conservative side \emph{banking housing} and \emph{monetary
  policy}. For Topic~11, which is about climate change/public health,
the term \emph{climate change} is used three times more often by a
speaker with ideal point 1 compared to one with $-1$. GST also detect
climate change as a partisan phrase with 94 (D) and 23 (R)
appearances per 100,000 phrases.

\section{Conclusion and Discussion}\label{sec:conc}

This study replicates GST in a wide sense through an
  alternative modeling approach, analyzing U.S.\ Senate speeches from
  1981 to 2017.  The differences of our approach compared to GST are
threefold: Firstly, TV-TBIP combines the class of topic models with
ideal point models. Thus, researchers do not have to manually specify
topics a-priori using key terms. Secondly, our approach is
unsupervised. This implies it can also be used to analyze data for
which no information about author positions is known hitherto as
pre-labeling is not required (party membership). Thirdly, our model is
able to detect the accordance of political parties on topic level
through polarity- and topic-specific term distributions in a
time-dynamic way.

Our results highlight a sharp rise in partisanship
  during the 1990s, which aligns with GST's conclusions. By inspecting
  individual ideal points, we compare Senators' positions on a latent
  scale. We also observe shifts in topic content over time. The
  climate change/public health topic emerges as a significant driver
  of partisanship, showing high term discordance between parties,
  while the natural resources and energy topic also differentiates
  between speakers across party lines. In contrast, some topics, like
  war veterans, display similar term compositions across both
  parties.

This sets the scene for further research which could
  include, e.g., alternative ways of modeling time variation in this
  model framework \citep[see][]{vavra2024a} or the
  incorporation of speaker-specific covariates that influence
  ideological positions \citep[see][]{vavra2024b}.

  \section*{Acknowledgments}

  The authors want to thank Keyon Vafa for many helpful comments and
  feedback.  This research was supported by the OeNB Jubiläumsfonds
  (project number: 18718).
 
\bibliography{arxiv}
\newpage
\appendix
\section{More Details on the TV-TBIP Model}\label{sec:more-deta-spec}
\subsection{Variational Inference}

Performing inference for a Bayesian model such as ours
consists of determining the posterior distribution of the parameters
and latent variables. Analytical solutions for the posterior
distribution are usually only available for very simple
models. Alternative approaches rely on computational methods to
approximate the posterior distribution. Markov chain Monte Carlo
methods enable to sample from the posterior by constructing a Markov
chain where the stationary distribution corresponds to the posterior
distribution. Due to the computational challenges of Markov chain
Monte Carlo methods when used for large latent variable models,
variational inference has emerged as the preferred method to
approximate the posterior for such models.

Variational inference is a technique that solve the estimation problem
by approximating the true posterior distribution with a simpler, more
tractable distribution.  Instead of relying on sampling such as Markov
chain Monte Carlo methods, variational inference recasts the inference
problem as an optimization problem. The core aspect in variational
inference consists of specifying a family of approximate densities
$\mathbb{Q}$ for the latent variables. Variational inference aims at
determining the density within this family that minimizes the
Kullback-Leibler (KL) divergence to the exact posterior.  In the
following, to simplify notation for describing variational inference,
we drop any time indices of our model. Let
$q_{\bm{\phi}}(\bm{\theta}, \bm{\beta}, \bm{\eta}, \bm{x}) \in
\mathbb{Q}$ be an approximate density with parameters $\bm{\phi}$ for
the true posterior density
$p(\bm{\theta}, \bm{\beta}, \bm{\eta}, \bm{x} | \bm{y})$, with
$\bm{\theta}, \bm{\beta}, \bm{\eta}, \bm{x}$ denoting parameters and
latent variables, and $\bm{y}$ represent the observed data, e.g.,
consisting of the frequency counts $\bm{c}$ and the speaker
information $\bm{s}$. In the following to further simplify notation,
we use $\bm{\vartheta}$ to denote the set of all parameters and latent
variables, i.e.,
$\bm{\vartheta}=\{\bm{\theta}, \bm{\beta}, \bm{\eta}, \bm{x}\}$.

The goal is to find the best approximation
$q_{\bm{\phi}^*}(\bm{\vartheta})$ by optimizing the parameters
$\bm{\phi}$ of the approximate density. This is typically done by
minimizing the Kullback-Leibler (KL) divergence between
$q_{\bm{\phi}}(\bm{\vartheta})$ and
$p(\bm{\vartheta}|\bm{y})$. Formally, the objective function for
variational inference can be expressed as:
\begin{displaymath}
  \underset{\bm{\phi}}{\arg\min}\, \text{KL}(q_{\bm{\phi}}(\bm{\vartheta}) \| p(\bm{\vartheta}|\bm{y})),
\end{displaymath}
where $\text{KL}(\cdot\|\cdot)$ denotes the Kullback-Leibler
divergence which may also be written as
\begin{displaymath}
  \text{KL}(q_{\bm{\phi}}(\bm{\vartheta}) \| p(\bm{\vartheta}|\bm{y})) =
  \mathbb{E}_{q_{\bm{\phi}}(\bm{\vartheta})}(\log(q_{\bm{\phi}}(\bm{\vartheta})))-\mathbb{E}_{q_{\bm{\phi}}(\bm{\vartheta})}(\log(p(\bm{\vartheta},\bm{y}))) + \log(p(\bm{y})).
\end{displaymath}
The evidence
$p(\bm{y})= \int p(\bm{\vartheta},\bm{y}) d\bm{\vartheta}$ is usually
unavailable in closed form.  However, the evidence also does not
depend on $\bm{\phi}$. Dropping the log evidence and taking the
negative results in the Evidence Lower Bound (ELBO) which can thus be
maximized to minimize the KL divergence:
\begin{displaymath}
  \text{ELBO}(\bm{\phi}) = \mathbb{E}_{q_{\bm{\phi}}(\bm{\vartheta})}[\log p(\bm{\vartheta},\bm{y})] - \mathbb{E}_{q_{\bm{\phi}}(\bm{\vartheta})}[\log q_{\bm{\phi}}(\bm{\vartheta})].
\end{displaymath}

For more details we refer to \cite{Blei+Kucukelbir+McAuliffe:2017}.
\subsection{Variational Families for TV-TBIP}
The variational family needs to provide sufficient flexibility to be
able to closely approximate the posterior.  For estimating the TV-TBIP
model, a mean-field variational family is used.  Let
$q_{\bm{\phi}^t} (\bm{\theta}^t, \bm{\beta}^t, \bm{\eta}^t, \bm{x}^t)$
be the variational family for time point $t$, indexed by variational
parameters $\bm{\phi}^t$.  Using a mean-field variational family
implies that the variational family factorizes over the latent
variables, where $i$ indexes speeches, $k$ indexes topics, and $s$
indexes authors:
\begin{align*}
  q_{\bm{\phi}^t} (\bm{\theta}^t, \bm{\beta}^t, \bm{\eta}^t, \bm{x}^t) &=
  \prod_{i} q(\bm{\theta}^t_i) \prod_k q(\bm{\beta}^t_k) \prod_k q(\bm{\eta}^t_k) \prod_s q(\bm{x}^t_s).
\end{align*}

We use log-normal variational distributions if the variables have
positive support and Gaussian variational distributions for variables
with unrestricted support:
\begin{align*}
{\theta}^t_{ik} &\sim \text{LogNormal} (\mu_{{\theta}^t_{ik}}, \sigma_{{\theta}^t_{ik}}^2),&
{\beta}^t_{kv} &\sim \text{LogNormal} (\mu_{\beta^t_{kv}}, \sigma_{\beta^t_{kv}}^2),\\
\eta^t_{kv} &\sim\text{Normal} (\mu_{\eta^t_{kv}}, \sigma_{\eta^t_{kv}}^2),&
x^t_s &\sim \text{Normal}(\mu_{x^t_s}, \sigma_{x^t_s}^2).
\end{align*}

The complete set of variational parameters is thus given by
$\bm{\phi}^t = \{\{\mu_{\theta^t_i}\}_i$,
$\{\sigma_{\theta^t_i}^2\}_i$, $\{\mu_{\beta^t_k}\}_k$,
$\{\sigma_{\beta^t_k}^2\}_k$,
$\{\mu_{\eta^t_k}\}_k$, $\{\sigma_{\eta^t_k}^2\}_k$,
$\{\mu_{x^t_s}\}_s$, $\{\sigma_{x^t_s}^2\}_s \}$.
These variational parameters are determined by maximizing the Evidence
Lower BOund (ELBO) which is equivalent to minimizing the
Kullback-Leibler divergence:
\begin{align*}
  \mathbb{E}_{q_{\bm{\phi}^t}} [\log p (\bm{\theta}^t, \bm{\beta}^t, \bm{\eta}^t, \bm{x}^t) +
  \log p (\bm{c}^t, \bm{s}^t | \bm{\theta}^t, \bm{\beta}^t, \bm{\eta}^t, \bm{x}^t) - \log q_{\bm{\phi}^t} (\bm{\theta}^t, \bm{\beta}^t, \bm{\eta}^t, \bm{x}^t)].
\end{align*}

Based on the estimated variational parameters, point estimates of the
model parameters and latent variables are obtained using the posterior
means induced by the estimated $\bm{\phi}^t$ values. The maximization
of the ELBO is performed using a general purpose optimizer, where
initial values need to be provided. More details are given in
\citet{Vafa_etal_2020}.

\subsection{Computational Details}
The analysis is performed using Python 3.7 \citep{python}, Tensorflow
1.15 GPU \citep{tensorflow2015-whitepaper} and scikit-learn 1.0.2
\citep{scikit-learn}. Running the optimizer with GPU support
considerably reduces runtime.  For optimizing the ELBO function we
make use of the Adam optimizer \citep{Kingma+Ba:2015} using 300,000
iterations per session. Inspection of the ELBO values indicated
convergence after about 100,000 iterations.  

\section{More Details on Data and Analysis}\label{sec:more-details-data}
\subsection{Data Description and Pre-processing} 
The Stanford University Social Science Data Collection database
\citep{Gentzkow+Shapiro+Taddy:2018} provides already processed text
data from the United States Congressional Record. The speeches given
by members of the U.S.\ Congress have been parsed and are stored as
transcripts of the full-text speeches together with some related
metadata information such as details about the speaker. We use the
``daily edition'' of the Senate speeches in the database for our
analysis which covers sessions~97 to 114 (1981--2017) and provides
data on the speech level.

Following GST and \citet{Vafa_etal_2020}, a number
of pre-processing steps were performed to obtain session-specific
document-term matrices with an aligned vocabulary from the files
containing the transcripted full-text speeches.
First, we removed punctuation and numbers, changed the text to
lower-case and eliminated stop words.
For tokenization, we followed GST and also used
bigrams. \citet{Gentzkow_etal_2019} argue that in certain applications
such as the analysis of partisan speech, single words are insufficient
to capture all important aspects. They point out that at least bigrams
are required to capture a limited amount of the dependence between
words and they also emphasize that bigrams are better able to gather
overtones and ideological phrases.  In addition, using bigrams instead
of single terms enhances interpretability of the resulting
topics when inspecting the topic-specific term prevalences.

For each session, we considered speeches given by Senators as well as
external speakers, such as House Representatives. We removed
speakers who gave less than 24 speeches in a particular session as
well as bigrams which were used by less than 10 speakers in a
particular session \citep[cf.][]{Vafa_etal_2020}. Finally, speeches
resulting in an empty frequency count, i.e., an empty row in the
document-term matrix, were omitted from further analysis. The complete
vocabulary spanning all the sessions from session~97 to 114 resulting
from these pre-processing steps consists of 12,527 unique bigrams.

Prior to these pre-processing steps, the dataset of the Senate
speeches contains 1,262,273 speeches given by 1,142 unique speakers.
After pre-processing we are left with 614,613 speeches given by 355
unique speakers (including Senators and external speakers) over a
period of 18 sessions, i.e., during the years 1981 until 2017.
Table~\ref{tab:data} displays summary statistics of the original data
before pre-processing as well as the data obtained after
pre-processing which was then used for estimating the TV-TBIP model.

\begin{table}[t!]
  \centering
  \begin{tabular}{ rrrrrrrrr}
    \hline
  &\multicolumn{2}{c}{Speakers} &\multicolumn{2}{c}{Senators}& \multicolumn{2}{c}{Speeches}&\multicolumn{2}{c}{$\varnothing$ Speeches}\\
  \hline
 Session & before & after & before & after& before & after & before & after\\
 \hline
      97 &         376 &         118 &  101 & 100&    110980 &       47088 &  295.16 &  399.05 \\
      98 &         347 &         121 &  101 &101&    97755 &       42608 &  281.71 &  352.13 \\
      99 &         304 &         104 &   101 &101&   107774 &       49271 &  354.52 &  473.76 \\
     100 &         302 &         105 &  101 &100 &   108618 &       49707 &  359.66 &  473.40 \\
     101 &         366 &         103 &  103 &101&    90024 &       44247 &  245.97 &  429.58 \\
     102 &         334 &         103 &  103 &101&    82233 &       42234 &  246.21 &  410.04 \\
     103 &         346 &         103 &  105&101&    83742 &       40957 &  242.03 &  397.64 \\
     104 &         244 &         103 &   102 &101&    89763 &       42879 &  367.88 &  416.30 \\
     105 &         191 &         100 &   100 &100&   66512 &       33591 &  348.23 &  335.91 \\
     106 &         200 &         101 &  101 &99&    66855 &       33833 &  334.28 &  334.98 \\
     107 &         212 &         100 &  101  &99&   62267 &       30895 &  293.71 &  308.95 \\
     108 &         150 &         100 &  100  &99&   65776 &       33571 &  438.51 &  335.71 \\
     109 &         170 &         101 &   101 &100&   53404 &       27842 &  314.14 &  275.66 \\
     110 &         151 &         102 &   102&100&    51919 &       27385 &  343.83 &  268.48 \\
     111 &         208 &         104 &  109&100&    37728 &       20236 &  181.38 &  194.58 \\
     112 &         126 &         100 &  101 &97&    32924 &       18421 &  261.30 &  184.21 \\
     113 &         114 &         100 &  105 &98&    29721 &       16249 &  260.71 &  162.49 \\
     114 &         111 &          99 &  100 &97&    24278 &       13599 &  218.72 &  137.36 \\
 \hline
\end{tabular}
\caption{Number of speakers (including Senators and external
  speakers), number of Senators, number of speeches and average number
  of speeches per speaker for each session before and after
  pre-processing.}
\label{tab:data}
\end{table}

\subsection{Practical Implementation of Model Specification and Estimation}

Following GST, we set the number of topics equal to 25 and specify
this number to be equal across sessions.  We want to emphasize that
there is no ``true'' number of topics \citep[see,
e.g.,][]{Roberts:2019}. Estimating only few topics implies that
topical content is very ``wide'', while a large number of topics might
result in very granular and maybe overlapping topics.  In contrast to
GST, we allow the topical content to change over time with the term
distributions of the topics being allowed to be session-specific.

To complete the model specification, we also have to fix the
parameters of the prior distributions of the topic prevalences
$\bm{\theta}^t$, the term prevalences $\bm{\beta}^t$ as well as the
polarity scores $\bm{\eta}^t$ and the ideal points $\bm{x}^t$. We use
standard normal priors for $\bm{\eta}^t$ and the ideal points
$\bm{x}^t$.  Both the topic prevalences as well as the topic-specific
term prevalences follow a-priori Gamma priors. Selecting the
parameters of these Gamma priors is crucial to induce sparsity in the
topic distributions of the speeches as well as the term distributions
of the topics. We follow \citet{Vafa_etal_2020} for the parameter
settings of these priors: We use the same parameter values for both
priors, i.e., $\alpha_1 = \gamma_1$ and $\alpha_2 = \gamma_2$. In
addition we use the same parameter values for shape and rate of the
Gamma distribution, which implies that the prior mean is equal to
1. The specific values selected for the parameters are
$\alpha_1 = \alpha_2 = 0.3$, i.e., a value smaller than one is used
which is crucial for sparsity and which implies a prior variance of 1
/ 0.3.

\subsection{Assessing Model Estimation in a Simulation Study}\label{sec:simulation}

We performed a simulation study to assess the performance of the
proposed model estimation scheme. We exploited the fact that TV-TBIP
model is a generative model and used the posterior mean estimates of
the parameters obtained when applying the estimation scheme to the
U.S.~Senate data from sessions 97--114 to draw new document-term
matrices for each session. To avoid extreme rates for the counts, the
estimated topic polarity scores were winsorized to be within $-1$ and
$+1$.

\begin{figure}
  \centering
\begin{minipage}{.48\textwidth}
  \centering
   \includegraphics[width=\textwidth, trim = 0 0 0 0, clip]{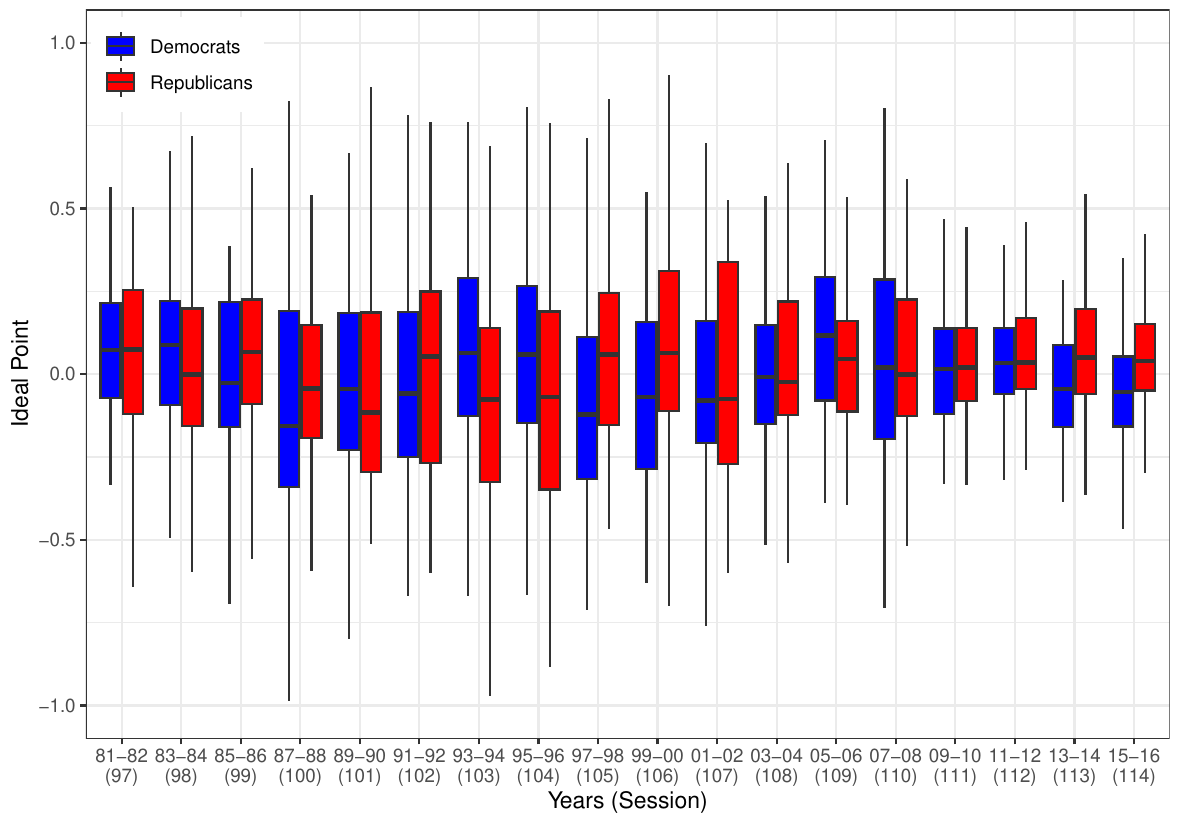}
\end{minipage}
\begin{minipage}{.48\textwidth}
   \centering
   \includegraphics[width=\textwidth, trim = 0 0 0 0, clip]{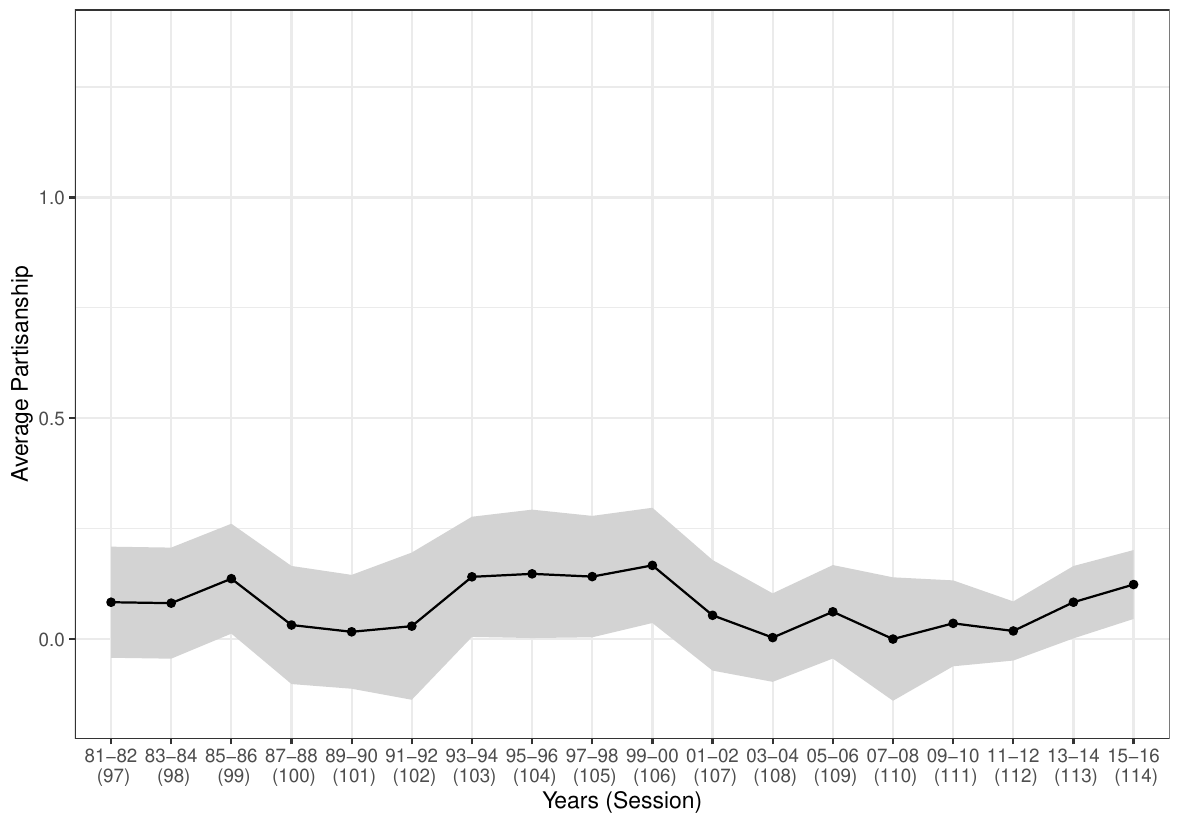}
\end{minipage}
\begin{minipage}{.48\textwidth}
  \centering
   \includegraphics[width=\textwidth, trim = 0 0 0 0, clip]{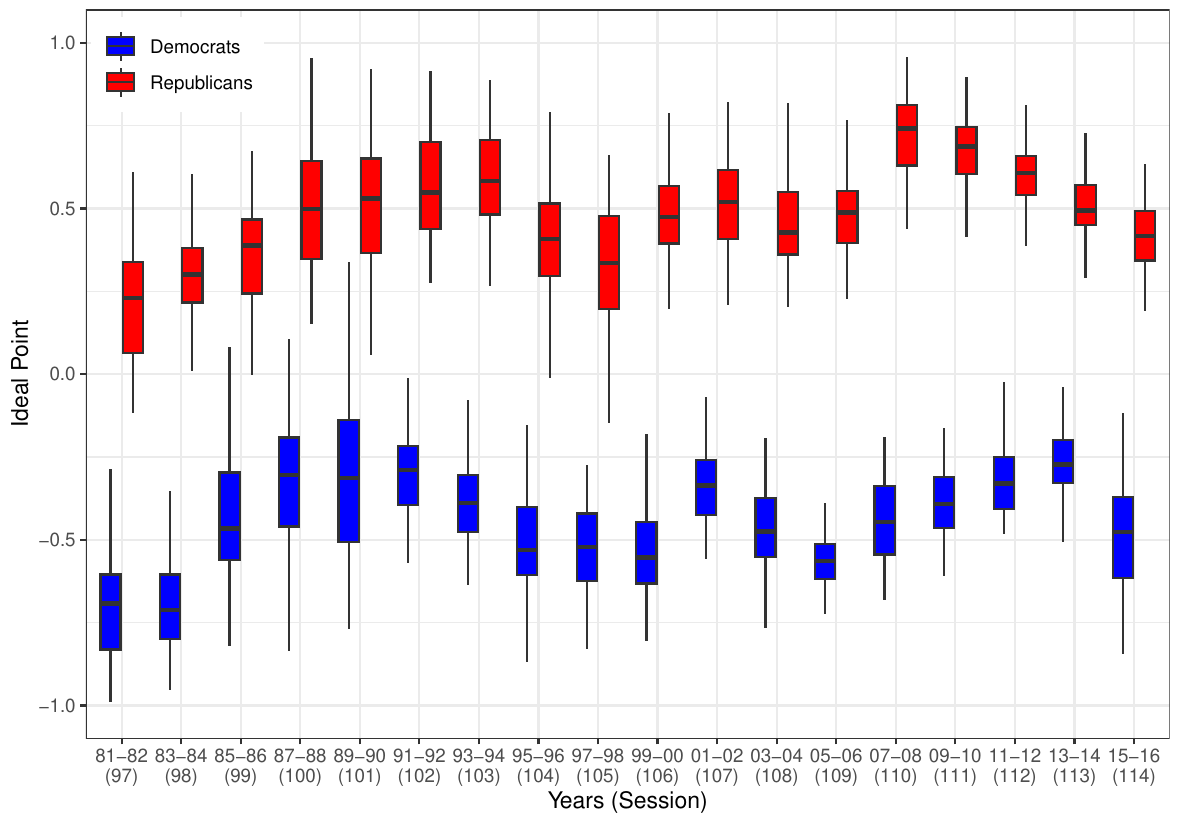}
\end{minipage}
\begin{minipage}{.48\textwidth}
   \centering
   \includegraphics[width=\textwidth, trim = 0 0 0 0, clip]{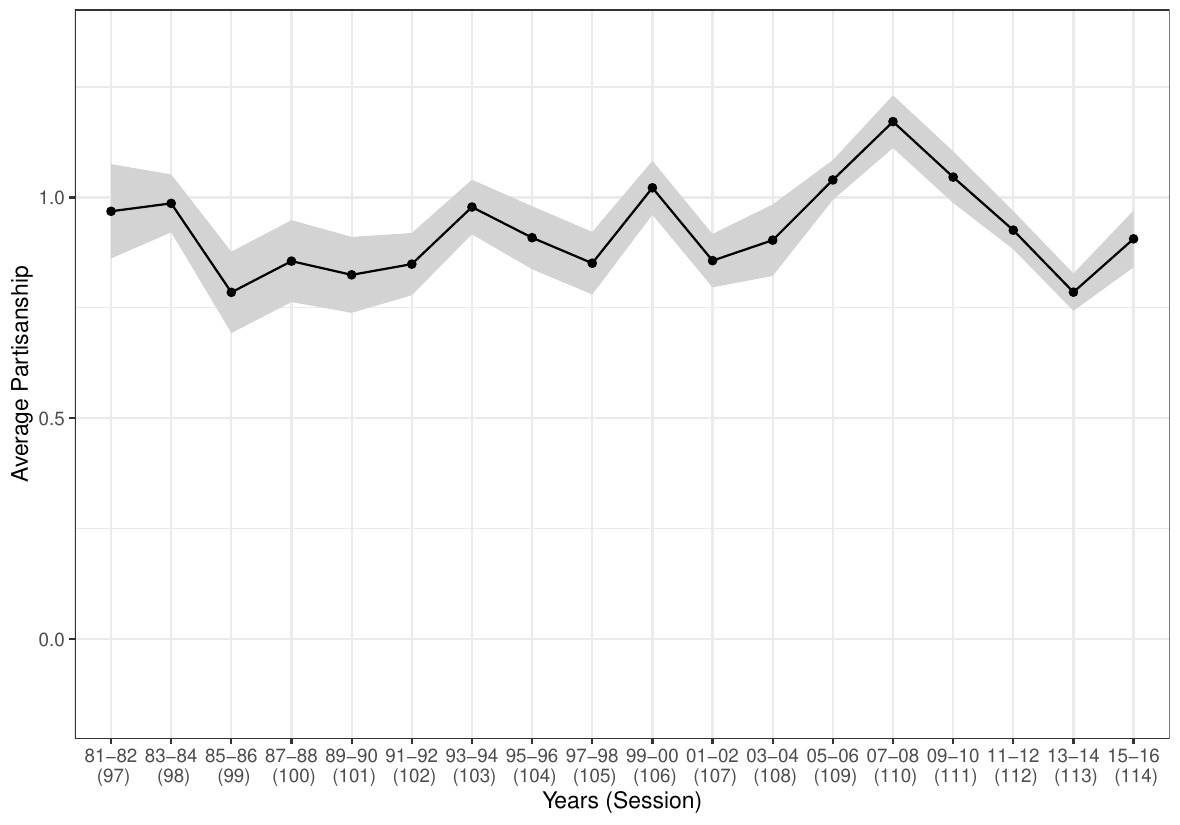}
\end{minipage}
\begin{minipage}{.48\textwidth}
  \centering
   \includegraphics[width=\textwidth, trim = 0 0 0 0, clip]{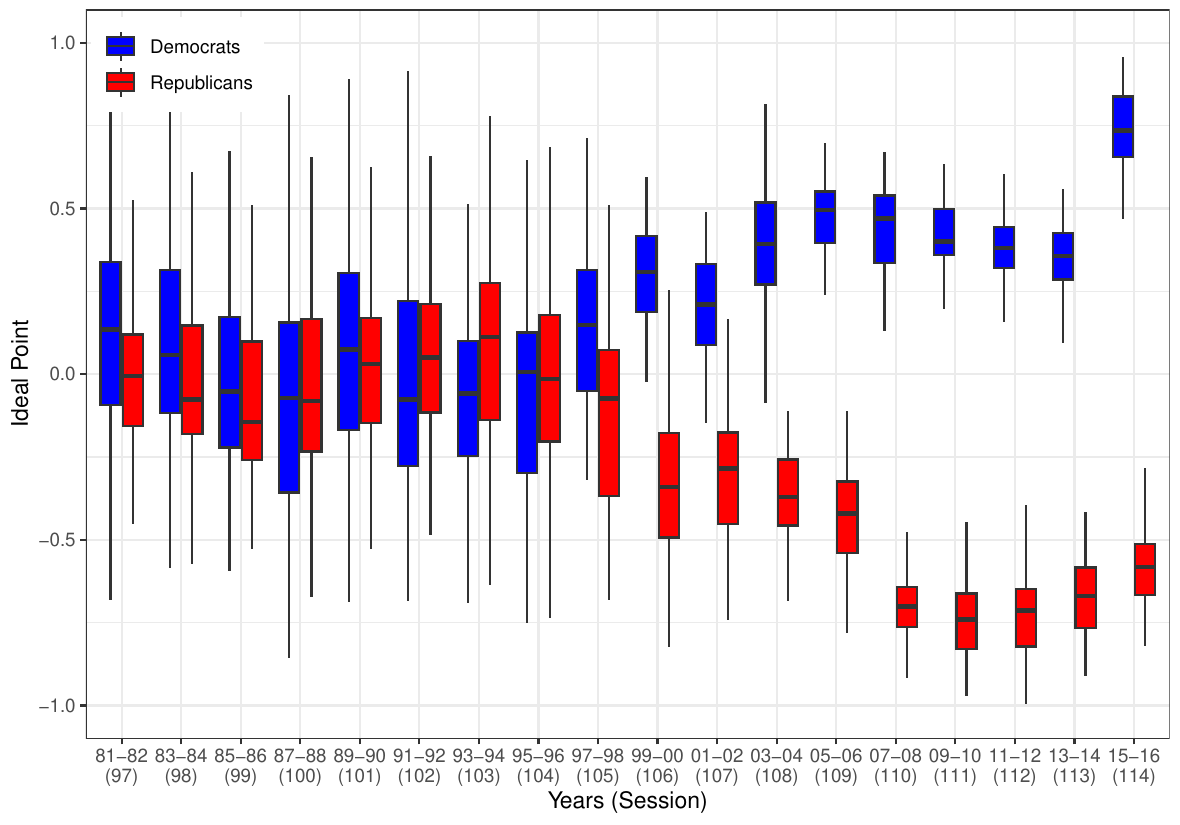}
\end{minipage}
\begin{minipage}{.48\textwidth}
   \centering
   \includegraphics[width=\textwidth, trim = 0 0 0 0, clip]{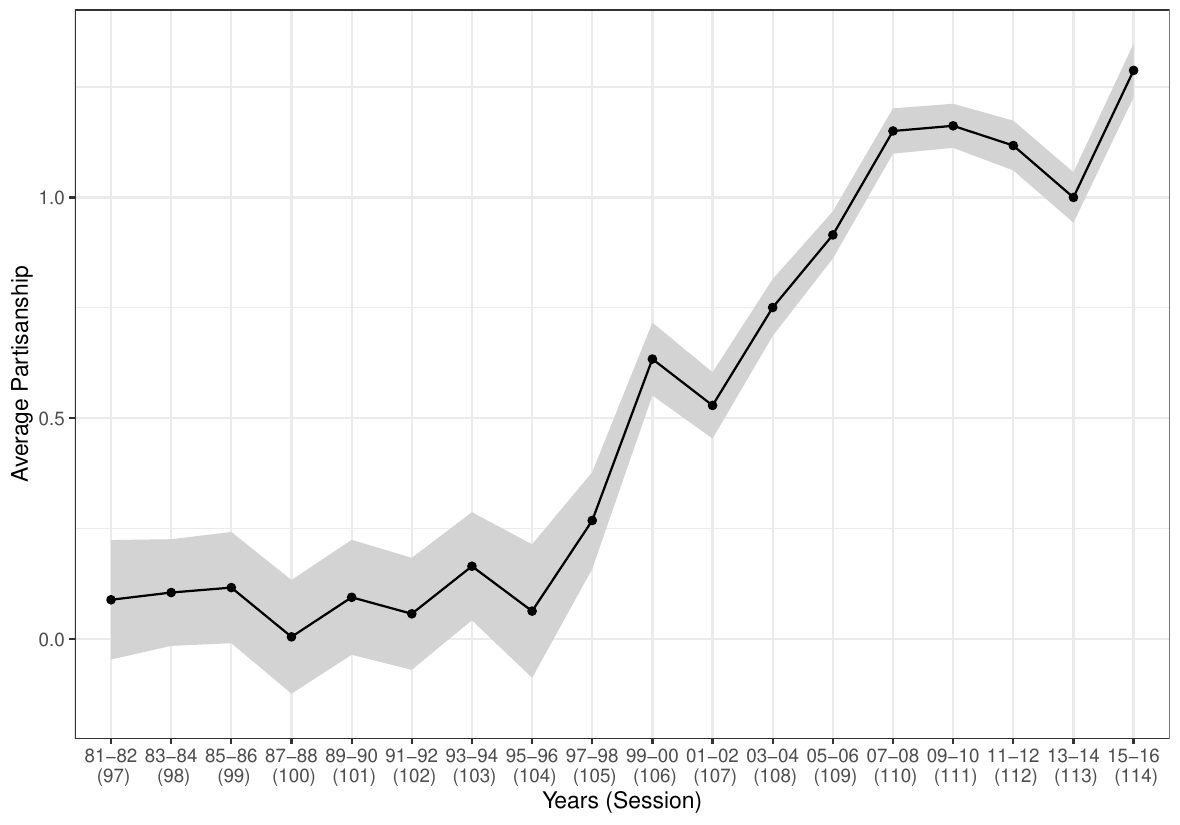}
\end{minipage}
\begin{minipage}{.48\textwidth}
  \centering
   \includegraphics[width=\textwidth, trim = 0 0 0 0, clip]{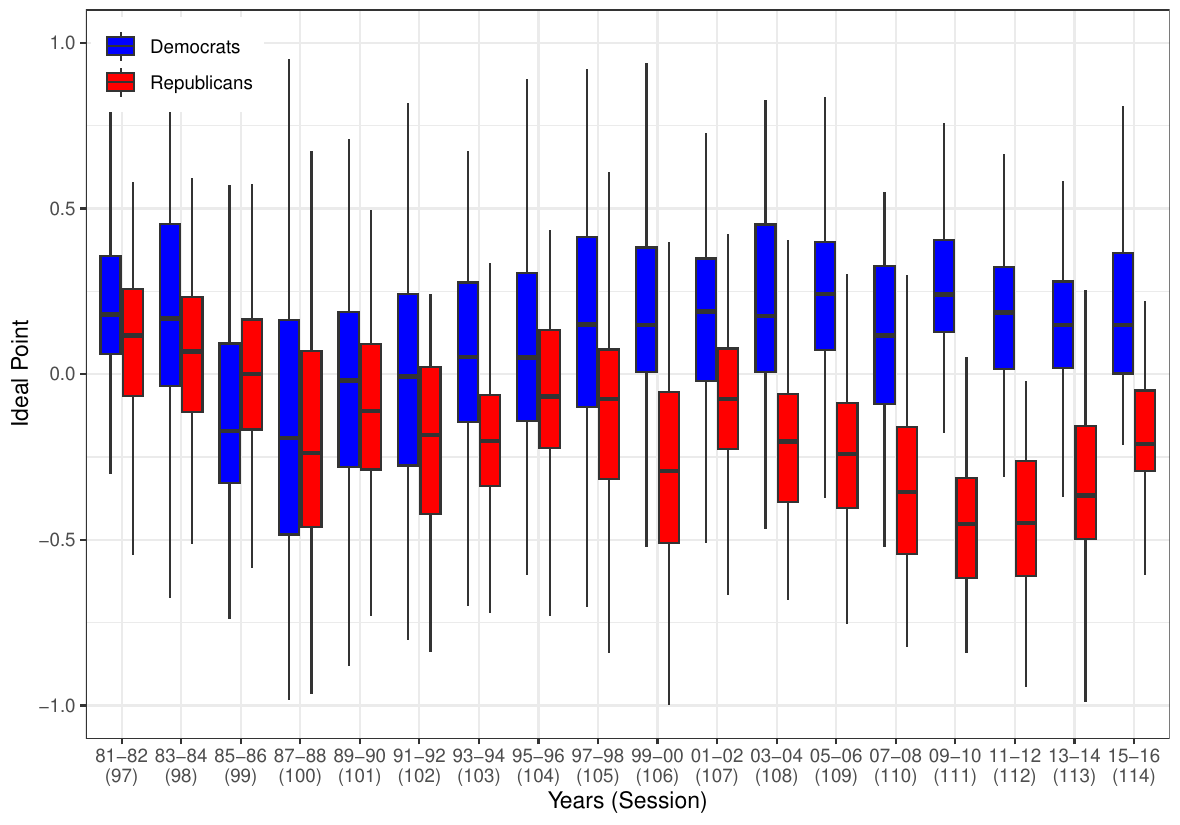}
\end{minipage}
\begin{minipage}{.48\textwidth}
   \centering
   \includegraphics[width=\textwidth, trim = 0 0 0 0, clip]{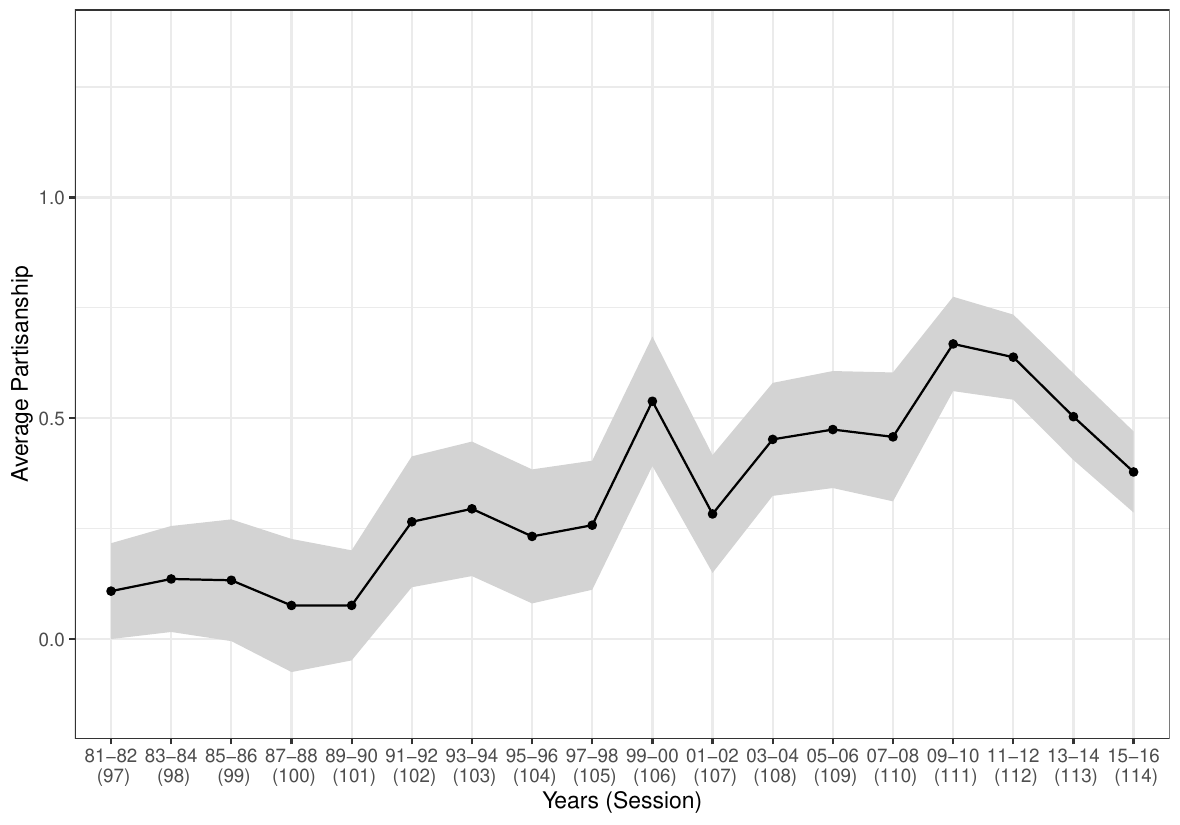}
\end{minipage}
\caption{From top to bottom: no party differences, fixed party
  differences, increasing party differences, estimated party
  differences. For each scenario: party- and session-specific ideal
  point distributions represented as box-plots. Estimated average
  partisanship over the years together with approximate pointwise 95\%
  confidence intervals (right). }
\label{fig:simulation}
\end{figure}

We considered four different scenarios which varied in the
distribution of the ideal points used across the Democrat and
Republican speakers:
\begin{enumerate}[(1)]
\item \emph{No party differences:} The ideal points were identical to
  zero for all speakers across all sessions.
\item \emph{Fixed party differences:} The ideal points of Democrat
  speakers were fixed to $-0.5$ and those of the Republican speakers
  to $+0.5$ for all speakers of the party across all sessions.
\item \emph{Increasing party differences:} The ideal points of
  Democrat and Republican speakers were set to zero for all speakers
  until session 100 and starting from session 101 gradually diverged by
  a linear decrease of 0.05 for each session for Democrats and a
  linear increase of 0.05 for each session for Republicans inducing an
  ideal point of $-0.7$ for Democrats and of $+0.7$ for Republicans in
  the last session.
\item \emph{Estimated party differences:} The estimated ideal points
  for the Democrat and Republican speakers in the U.S.~Senate for each
  session were used.
\end{enumerate}
We used the same estimation method as for the empirical application
on the U.S.~Senate data. The results are obtained using a different
computing environment than for the empirical application and are based
on Python 3.6.13, Tensorflow 1.15 GPU and scikit-learn 0.24.2. The
initialization was robustified by winsorizing the estimated polarity
scores from the previous session to $-1$ and $+1$.

We performed the same analysis of the results including the
visualization of the estimated party- and session-specific ideal point
distributions represented as box-plots and the estimated average
partisanship over the years together with approximate pointwise 95\%
confidence intervals. The results for the four scenarios are shown in
Figure~\ref{fig:simulation} after re-scaling the newly estimated
polarity scores to have the same robust standard deviation (determined
based on the inter-quartile range) than those obtained for the
original data. The estimation method is able to differentiate well
between the scenarios and to identify that there are no party-specific
differences in the first scenario, clear differences between the two
parties in the second scenario and no difference at the beginning with
an increasing difference over session in the third scenario. The
results for the forth scenario indicate that the procedure is able to
reliably re-estimate the ideal point distributions and the average
partisanship.

\section{Additional Results}\label{sec:additional-results}

\subsection{Party-Specific Ideal Point Distributions Across Sessions}
\begin{figure}[t!]
  \centering
\includegraphics[width=\textwidth]{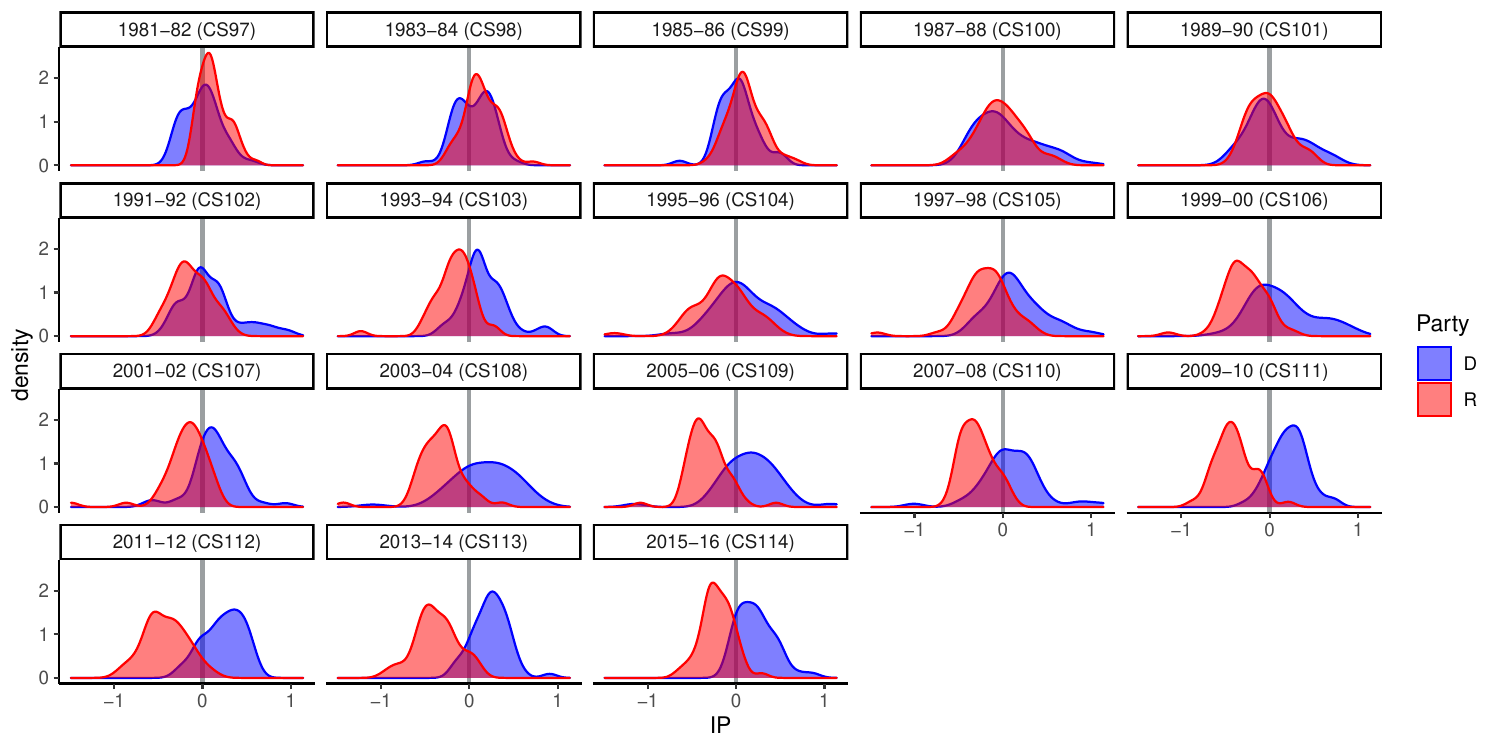}
\caption{Kernel density plots characterizing session- and party-specific ideal point distributions. 
  \label{fig:IP2}}
\end{figure}

The session- and party-specific ideal point distributions estimated by
TV-TBIP are analyzed using box-plots in the main paper.  An
alternative view is provided in Figure~\ref{fig:IP2} where their
kernel density estimates are visualized. The density estimates for
each of the two parties for the same session are combined in one panel
and the sessions are arranged row-wise across time. Kernel density
estimates provide a more flexible and detailed view on the
distribution of the ideal points compared to the box-plots. The kernel
density estimates indicate that the within party- and session-specific
ideal point distributions are approximately unimodal and that the
modes between the two parties separate over time, reducing also
considerably the overlap of the estimated densities.

\subsection{Comparison to DW-Nominate Scores}
We determine standardized average partisanship estimates across time
using the first dimension of the DW-Nominate scores and compare them
with the text-based average partisanship estimates from the TV-TBIP
model.  The session-specific average partisanship based on the
DW-Nominate scores is obtained as the difference between the average
DW-Nominate scores of Republicans and Democrats for each
session.\footnote{The data were downloaded from \url{https://votev
    iew.com/data} using Data Type: Congressional Parties, Chamber:
  Senate Only, Congress: All. with the variable nominate-dim1-mean
  used for analysis.}

\begin{figure}[b!]
\centering
\includegraphics[width=0.8\textwidth]{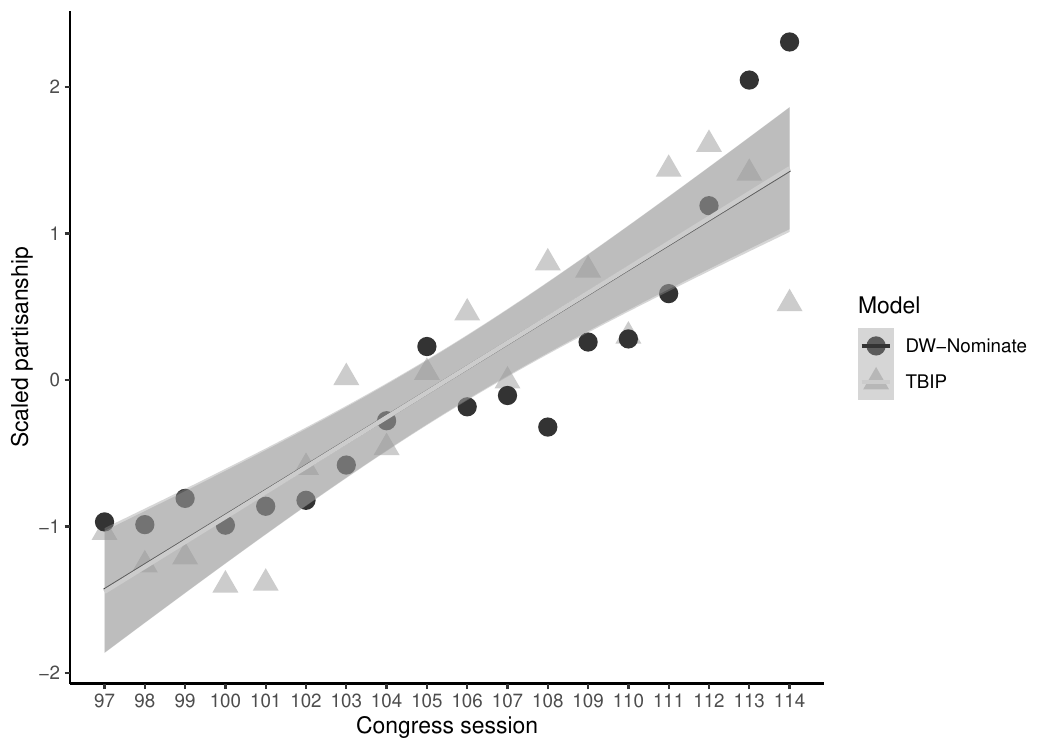}
\caption{Standardized aggregate partisanship estimates on party level
  based on DW-Nominate scores and the TV-TBIP ideal points across time
  together with fitted linear regression lines and corresponding
  95\% confidence intervals for the means. The correlation between
  these partisanship estimates is $0.77$.
  \label{fig:IPvox}}
\end{figure}

Figure~\ref{fig:IPvox} provides scatter plots of the sessions on the
$x$-axis versus the standardized average partisanship estimates for
the TV-TBIP model as well as the DW-Nominate scores on the
$y$-axis. Clearly both measures exhibit a similar increase over
time. This is also indicated by the overlap of the fitted regression
lines and their 95\% confidence intervals for the mean which are also
included in the plot. This implies that our text-based average
partisanship measure captures essentially the same effect over time as
the DW-Nominate scores.

\subsection{Ideological Positions on Speaker Level}

\begin{figure}[b!]
  \centering
  \includegraphics[width=\textwidth]{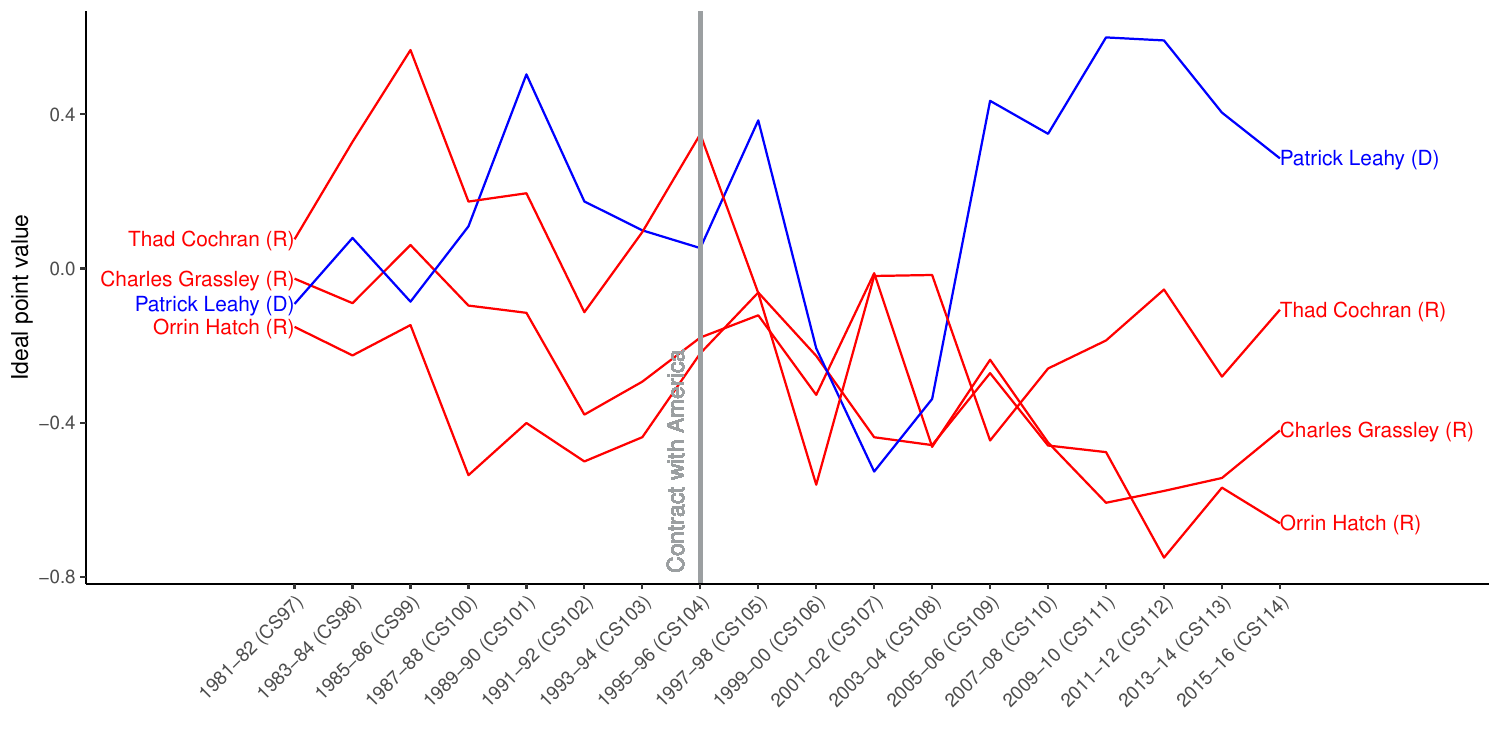}
\caption{Development of estimated ideal points over time of those four
  Senators who were members of the Senate from session~97 to 114.}
  \label{fig:IPcomp}
\end{figure}

We further investigate the ideal points estimated by the TV-TBIP for
the speakers and in particular their evolvement over time by focusing
on speakers who were members of the Senate during the whole analysis
period and hence have an ideal point estimated for each
session. Figure~\ref{fig:IPcomp} displays the development of estimated
ideal points of the four speakers who were members of the Senate
during the whole analysis period.  These speakers are three
Republicans and one Democrat, namely Thad Cochran (R), Charles
Grassley (R), Orrin Hatch (R) and Patrick Leahy (D).  Inspection of
the evolvement of the ideal points for those four speakers provides
the following interesting insights: Firstly, we can discern a drop in
the ideal points of all four Senators after the \emph{Contract with
  America}.  The reasons are not obvious and one can only
speculate. One possible explanation could be the Congress turnover
which took place. At this point in time, it was the first time after
40 years that Republicans had the majority in the Congress.  Secondly,
Figure~\ref{fig:IPcomp} confirms that the ideological positions
inferred by TV-TBIP are reasonable. Among these four Senators, the
most liberal Senator Patrick Leahy (D) clearly has the highest ideal
point values for the latter time periods, while the three Republican
Senators are after the Contract with America consistently positioned
on the negative side of the ideological scale. Among the three
Republicans displayed in Figure~\ref{fig:IPcomp}, we find Thad Cochran
to be the most liberal one. Indeed Thad Cochran is usually considered
to be more moderate than most of his Republican colleagues
\citep[cf.,][]{538}. For example, the New York Times arranged in 2017
Republican Senators based on ideology and reported that Thad Cochran
was the fourth most moderate Republican \citep[see][]{NYT}.  On the
other side of the spectrum we find Orrin Hatch (R), one of the leading
figures behind the Senate's anti-terrorism bill, and a person who is
strongly opposed to abortion \citep[see, e.g.,][]{wiki:orrin}.

\input{tableTBIP}

Table~\ref{tab:IPs} displays the estimated ideological positions of
selected Republican and Democrat speakers for the last 10 sessions,
i.e., we consider results from session~105 onwards. For Republicans as
well as Democrats, we display the five most liberal and conservative
speakers according to their mean ideal point over the last 10 sessions
per party.  Table~\ref{tab:IPs} indicates that according to TV-TBIP,
the most liberal Democrats are Byron Dorgan, Dale Bumpers, Thomas
Harkin, Christopher Murphy and Paul Wellstone.

As a Chairman of the Senate Energy Panel, Dorgan was
an early supporter of renewable energy, and in
Section~3.2 of the main manuscript we show that the energy topic is very
polarizing between the two parties.  We also find Paul Wellston among
the top five most liberal Democratic Senators. This is completely in
line with the DW-Nominate scores. According to these voting-based
scores, he is the most liberal Democratic Senator during
sessions~105--107.  Analyzing the conservative spectrum of Democratic
Senators in the Senate, we find that Evan Bayh is most
conservative. This is again in line with the DW-Nominate scores, where
he is ranked the third most conservative Democrat of session~106.

Proceeding with the Republican party, we find Jesse Helms, a Senator
from North Carolina, to be the most conservative Republican. The New
York Times \cite[see][]{Helms1} stated that Helms was ``bitterly
opposed'' to federal financing for research and treatment of AIDS
which he believed was God's punishment for homosexuals \citep[see,
e.g.,][]{Helms}. According to the DW-Nominate scores, Jesse Helms is
also ranked as the most conservative Republican for, e.g.,
sessions~106 and 107. Based on the mean ideal point value, the second
ranked among the most conservative Republicans is Gordon Smith. This
is in line with the following statement: ``Smith is often described as
politically moderate, but has strong conservative credentials as
well'' \citep[see][]{Smith}.

On the liberal side of the Republicans, TV-TBIP places the Californian
House Representative Howard McKeon, and Senators Phil Gramm, Dirk
Kempthorne, John Chafee and Ed Bryant. John Chafee is among the most
liberal Republicans according to the DW-Nominate scores which rate him
as being more liberal than $96\%$ of Republicans in session~105.

Finally, for independent speakers, the estimated ideal points meet
expectations. Bernie Sanders is the most liberal one, also in the
Democratic party he would be ranked second. On the other hand, former
Democrat Joe Liebermann is categorized as having a slightly liberal
position and former Republican Jim Jeffords seems to have a tendency
to be on the conservative side.

\end{document}

%% file: tableneutral.tex
\begin{table}

  \centering
\resizebox{0.99\textwidth}{!}{%
  \begin{tabular}{clcc}
  \hline
 \#&Topic & Session 97 (from January 3, 1981, to January 3, 1983) & Session 114 (from January 3, 2015, to January 3, 2017) \\ 
  \hline
1&Trade& $\underset{(0.315)}{\textrm{united states}}$, $\underset{(0.008)}{	\textrm{foreign policy}}$, $\underset{(0.007)}{	\textrm{international trade}}$, $\underset{(0.006)}{	\textrm{foreign relations}}$, $\underset{(0.005)}{	\textrm{acid rain}}$ & $\underset{(0.389)}{	\textrm{united states}}$, $\underset{(0.022)}{	\textrm{puerto rico}}$, $\underset{(0.021)}{	\textrm{trade agreements}}$, $\underset{(0.014)}{	\textrm{trade promotion}}$, $\underset{(0.013)}{	\textrm{promotion authority}}$ \\ 
2&Export/Import& $\underset{(0.03)}{	\textrm{private sector}}$, $\underset{(0.021)}{	\textrm{small businesses}}$, $\underset{(0.019)}{	\textrm{tax credit}}$, $\underset{(0.016)}{	\textrm{internal revenue}}$, $\underset{(0.011)}{	\textrm{clean air}}$ & $\underset{(0.049)}{	\textrm{small businesses}}$, $\underset{(0.026)}{	\textrm{american workers}}$, $\underset{(0.025)}{	\textrm{exportimport bank}}$, $\underset{(0.02)}{	\textrm{exim bank}}$, $\underset{(0.019)}{	\textrm{economic growth}}$ \\ 
3&Commemoration& $\underset{(0.071)}{	\textrm{legal services}}$, $\underset{(0.032)}{	\textrm{services corporation}}$, $\underset{(0.007)}{	\textrm{legal assistance}}$, $\underset{(0.007)}{	\textrm{older americans}}$, $\underset{(0.006)}{	\textrm{human resources}}$ & $\underset{(0.024)}{	\textrm{medical research}}$, $\underset{(0.02)}{	\textrm{state university}}$, $\underset{(0.018)}{	\textrm{high school}}$, $\underset{(0.018)}{	\textrm{passed away}}$, $\underset{(0.013)}{	\textrm{national institutes}}$ \\ 
4&Supreme court& $\underset{(0.095)}{	\textrm{supreme court}}$, $\underset{(0.03)}{	\textrm{federal courts}}$, $\underset{(0.015)}{	\textrm{public schools}}$, $\underset{(0.015)}{	\textrm{federal court}}$, $\underset{(0.014)}{	\textrm{school prayer}}$ & $\underset{(0.146)}{	\textrm{supreme court}}$, $\underset{(0.03)}{	\textrm{attorney general}}$, $\underset{(0.017)}{	\textrm{justice scalia}}$, $\underset{(0.015)}{	\textrm{district court}}$, $\underset{(0.013)}{	\textrm{judicial nominees}}$ \\ 
5&Finances& $\underset{(0.13)}{	\textrm{interest rates}}$, $\underset{(0.048)}{	\textrm{high interest}}$, $\underset{(0.035)}{	\textrm{federal reserve}}$, $\underset{(0.021)}{	\textrm{interest rate}}$, $\underset{(0.016)}{	\textrm{reserve board}}$ & $\underset{(0.061)}{	\textrm{wall street}}$, $\underset{(0.02)}{	\textrm{federal reserve}}$, $\underset{(0.012)}{	\textrm{protection act}}$, $\underset{(0.011)}{	\textrm{financial crisis}}$, $\underset{(0.008)}{	\textrm{monetary policy}}$ \\ 
6&Transport& $\underset{(0.077)}{	\textrm{natural gas}}$, $\underset{(0.015)}{	\textrm{energy assistance}}$, $\underset{(0.011)}{	\textrm{gas prices}}$, $\underset{(0.01)}{	\textrm{policy act}}$, $\underset{(0.008)}{	\textrm{lowincome energy}}$ & $\underset{(0.086)}{	\textrm{home state}}$, $\underset{(0.021)}{	\textrm{important issue}}$, $\underset{(0.016)}{	\textrm{critically important}}$, $\underset{(0.012)}{	\textrm{transportation system}}$, $\underset{(0.01)}{	\textrm{surface transportation}}$ \\ 
7&Law enforcement& $\underset{(0.045)}{	\textrm{law enforcement}}$, $\underset{(0.018)}{	\textrm{criminal justice}}$, $\underset{(0.018)}{	\textrm{violent crime}}$, $\underset{(0.017)}{	\textrm{coast guard}}$, $\underset{(0.014)}{	\textrm{attorney general}}$ & $\underset{(0.082)}{	\textrm{law enforcement}}$, $\underset{(0.026)}{	\textrm{human trafficking}}$, $\underset{(0.021)}{	\textrm{mental health}}$, $\underset{(0.014)}{	\textrm{gun violence}}$, $\underset{(0.013)}{	\textrm{criminal justice}}$ \\ 
8&Middle East& $\underset{(0.078)}{	\textrm{saudi arabia}}$, $\underset{(0.068)}{	\textrm{middle east}}$, $\underset{(0.026)}{	\textrm{foreign policy}}$, $\underset{(0.018)}{	\textrm{persian gulf}}$, $\underset{(0.015)}{	\textrm{foreign relations}}$ & $\underset{(0.143)}{	\textrm{national security}}$, $\underset{(0.034)}{	\textrm{middle east}}$, $\underset{(0.03)}{	\textrm{united states}}$, $\underset{(0.016)}{	\textrm{al qaeda}}$, $\underset{(0.013)}{	\textrm{islamic state}}$ \\ 
9&Budget& $\underset{(0.071)}{	\textrm{balanced budget}}$, $\underset{(0.032)}{	\textrm{federal budget}}$, $\underset{(0.029)}{	\textrm{federal spending}}$, $\underset{(0.022)}{	\textrm{budget process}}$, $\underset{(0.017)}{	\textrm{debt limit}}$ & $\underset{(0.019)}{	\textrm{appropriations process}}$, $\underset{(0.018)}{	\textrm{balanced budget}}$, $\underset{(0.017)}{	\textrm{control act}}$, $\underset{(0.014)}{	\textrm{budget act}}$, $\underset{(0.013)}{	\textrm{budget control}}$ \\ 
10&Education& $\underset{(0.033)}{	\textrm{budget authority}}$, $\underset{(0.016)}{	\textrm{budget act}}$, $\underset{(0.011)}{	\textrm{supplemental appropriations}}$, $\underset{(0.01)}{	\textrm{block grant}}$, $\underset{(0.01)}{	\textrm{loan program}}$ & $\underset{(0.057)}{	\textrm{high school}}$, $\underset{(0.03)}{	\textrm{left behind}}$, $\underset{(0.025)}{	\textrm{child left}}$, $\underset{(0.023)}{	\textrm{higher education}}$, $\underset{(0.015)}{	\textrm{school districts}}$ \\ 
11&Climate change& $\underset{(0.045)}{	\textrm{nuclear waste}}$, $\underset{(0.024)}{	\textrm{spent fuel}}$, $\underset{(0.022)}{	\textrm{nuclear power}}$, $\underset{(0.018)}{	\textrm{clinch river}}$, $\underset{(0.015)}{	\textrm{public health}}$ & $\underset{(0.074)}{	\textrm{climate change}}$, $\underset{(0.045)}{	\textrm{public health}}$, $\underset{(0.03)}{	\textrm{clean water}}$, $\underset{(0.021)}{	\textrm{zika virus}}$, $\underset{(0.016)}{	\textrm{fossil fuel}}$ \\ 
12&Taxes& $\underset{(0.061)}{	\textrm{tax cut}}$, $\underset{(0.029)}{	\textrm{tax cuts}}$, $\underset{(0.015)}{	\textrm{tax increase}}$, $\underset{(0.014)}{	\textrm{economic recovery}}$, $\underset{(0.013)}{	\textrm{tax rates}}$ & $\underset{(0.057)}{	\textrm{middle class}}$, $\underset{(0.028)}{	\textrm{tax credit}}$, $\underset{(0.025)}{	\textrm{tax code}}$, $\underset{(0.02)}{	\textrm{republican budget}}$, $\underset{(0.018)}{	\textrm{tax reform}}$ \\ 
13&Civil rights& $\underset{(0.06)}{	\textrm{voting rights}}$, $\underset{(0.048)}{	\textrm{rights act}}$, $\underset{(0.036)}{	\textrm{civil rights}}$, $\underset{(0.02)}{	\textrm{justice department}}$, $\underset{(0.015)}{	\textrm{attorney general}}$ & $\underset{(0.046)}{	\textrm{civil rights}}$, $\underset{(0.022)}{	\textrm{voting rights}}$, $\underset{(0.021)}{	\textrm{equal pay}}$, $\underset{(0.018)}{	\textrm{rights act}}$, $\underset{(0.014)}{	\textrm{fair housing}}$ \\ 
14&Federal government& $\underset{(0.109)}{	\textrm{federal government}}$, $\underset{(0.022)}{	\textrm{local governments}}$, $\underset{(0.009)}{	\textrm{public law}}$, $\underset{(0.009)}{	\textrm{federal funds}}$, $\underset{(0.008)}{	\textrm{federal agencies}}$ & $\underset{(0.157)}{	\textrm{federal government}}$, $\underset{(0.037)}{	\textrm{private sector}}$, $\underset{(0.026)}{	\textrm{taxpayer dollars}}$, $\underset{(0.02)}{	\textrm{federal agencies}}$, $\underset{(0.015)}{	\textrm{local governments}}$ \\ 
15&Health care& $\underset{(0.057)}{	\textrm{health care}}$, $\underset{(0.016)}{	\textrm{veterans administration}}$, $\underset{(0.015)}{	\textrm{older americans}}$, $\underset{(0.014)}{	\textrm{medical care}}$, $\underset{(0.011)}{	\textrm{senior citizens}}$ & $\underset{(0.161)}{	\textrm{health care}}$, $\underset{(0.04)}{	\textrm{affordable care}}$, $\underset{(0.039)}{	\textrm{care act}}$, $\underset{(0.038)}{	\textrm{health insurance}}$, $\underset{(0.038)}{	\textrm{planned parenthood}}$ \\ 
16&Rhetoric& $\underset{(0.081)}{	\textrm{peace corps}}$, $\underset{(0.022)}{	\textrm{foreign relations}}$, $\underset{(0.009)}{	\textrm{distinguished colleague}}$, $\underset{(0.008)}{	\textrm{intelligence activities}}$, $\underset{(0.005)}{	\textrm{equally divided}}$ & $\underset{(0.046)}{	\textrm{next week}}$, $\underset{(0.045)}{	\textrm{equally divided}}$, $\underset{(0.035)}{	\textrm{en bloc}}$, $\underset{(0.025)}{	\textrm{important legislation}}$, $\underset{(0.017)}{	\textrm{final passage}}$ \\ 
17&Wars& $\underset{(0.136)}{	\textrm{soviet union}}$, $\underset{(0.026)}{	\textrm{world war}}$, $\underset{(0.02)}{	\textrm{foreign policy}}$, $\underset{(0.018)}{	\textrm{war ii}}$, $\underset{(0.013)}{	\textrm{chemical weapons}}$ & $\underset{(0.061)}{	\textrm{world war}}$, $\underset{(0.051)}{	\textrm{war ii}}$, $\underset{(0.045)}{	\textrm{puerto rico}}$, $\underset{(0.014)}{	\textrm{vietnam war}}$, $\underset{(0.013)}{	\textrm{loved ones}}$ \\ 
18&Nuclear arms& $\underset{(0.048)}{	\textrm{nuclear weapons}}$, $\underset{(0.045)}{	\textrm{arms control}}$, $\underset{(0.025)}{	\textrm{nuclear war}}$, $\underset{(0.022)}{	\textrm{foreign relations}}$, $\underset{(0.016)}{	\textrm{nuclear arms}}$ & $\underset{(0.037)}{	\textrm{united states}}$, $\underset{(0.035)}{	\textrm{nuclear weapons}}$, $\underset{(0.031)}{	\textrm{nuclear weapon}}$, $\underset{(0.026)}{	\textrm{nuclear program}}$, $\underset{(0.02)}{	\textrm{middle east}}$ \\ 
19&Social security& $\underset{(0.22)}{	\textrm{social security}}$, $\underset{(0.028)}{	\textrm{security system}}$, $\underset{(0.02)}{	\textrm{minimum benefit}}$, $\underset{(0.02)}{	\textrm{trust fund}}$, $\underset{(0.018)}{	\textrm{security benefits}}$ & $\underset{(0.142)}{	\textrm{social security}}$, $\underset{(0.056)}{	\textrm{trust fund}}$, $\underset{(0.023)}{	\textrm{highway trust}}$, $\underset{(0.011)}{	\textrm{older americans}}$, $\underset{(0.011)}{	\textrm{security administration}}$ \\ 
20&Army& $\underset{(0.059)}{	\textrm{armed services}}$, $\underset{(0.05)}{	\textrm{air force}}$, $\underset{(0.022)}{	\textrm{armed forces}}$, $\underset{(0.02)}{	\textrm{basing mode}}$, $\underset{(0.017)}{	\textrm{national guard}}$ & $\underset{(0.051)}{	\textrm{air force}}$, $\underset{(0.048)}{	\textrm{armed services}}$, $\underset{(0.045)}{	\textrm{defense authorization}}$, $\underset{(0.042)}{	\textrm{national defense}}$, $\underset{(0.023)}{	\textrm{national guard}}$ \\ 
21&Natural resources& $\underset{(0.025)}{	\textrm{oil companies}}$, $\underset{(0.022)}{	\textrm{natural resources}}$, $\underset{(0.017)}{	\textrm{windfall profit}}$, $\underset{(0.017)}{	\textrm{crude oil}}$, $\underset{(0.016)}{	\textrm{profit tax}}$ & $\underset{(0.021)}{	\textrm{natural resources}}$, $\underset{(0.02)}{	\textrm{keystone xl}}$, $\underset{(0.019)}{	\textrm{natural gas}}$, $\underset{(0.017)}{	\textrm{energy efficiency}}$, $\underset{(0.017)}{	\textrm{water conservation}}$ \\ 
22&Human rights& $\underset{(0.08)}{	\textrm{human rights}}$, $\underset{(0.03)}{	\textrm{united nations}}$, $\underset{(0.027)}{	\textrm{genocide convention}}$, $\underset{(0.02)}{	\textrm{foreign relations}}$, $\underset{(0.019)}{	\textrm{state department}}$ & $\underset{(0.056)}{	\textrm{human rights}}$, $\underset{(0.041)}{	\textrm{state department}}$, $\underset{(0.032)}{	\textrm{foreign relations}}$, $\underset{(0.023)}{	\textrm{religious freedom}}$, $\underset{(0.019)}{	\textrm{united nations}}$ \\ 
23&Security& $\underset{(0.051)}{	\textrm{national security}}$, $\underset{(0.033)}{	\textrm{inspector general}}$, $\underset{(0.023)}{	\textrm{defense department}}$, $\underset{(0.021)}{	\textrm{national defense}}$, $\underset{(0.014)}{	\textrm{defense spending}}$ & $\underset{(0.108)}{	\textrm{homeland security}}$, $\underset{(0.034)}{	\textrm{cyber security}}$, $\underset{(0.017)}{	\textrm{usa freedom}}$, $\underset{(0.017)}{	\textrm{freedom act}}$, $\underset{(0.015)}{	\textrm{inspector general}}$ \\ 
24&Puerto Rico& $\underset{(0.04)}{	\textrm{food stamp}}$, $\underset{(0.021)}{	\textrm{school lunch}}$, $\underset{(0.02)}{	\textrm{food stamps}}$, $\underset{(0.019)}{	\textrm{puerto rico}}$, $\underset{(0.019)}{	\textrm{stamp program}}$ & $\underset{(0.016)}{	\textrm{economic development}}$, $\underset{(0.014)}{	\textrm{local communities}}$, $\underset{(0.014)}{	\textrm{puerto rico}}$, $\underset{(0.011)}{	\textrm{foster care}}$, $\underset{(0.01)}{	\textrm{much needed}}$ \\ 
25&Party rhetoric& $\underset{(0.041)}{	\textrm{executive branch}}$, $\underset{(0.032)}{	\textrm{legislative veto}}$, $\underset{(0.024)}{	\textrm{regulatory reform}}$, $\underset{(0.018)}{	\textrm{federal trade}}$, $\underset{(0.017)}{	\textrm{trade commission}}$ & $\underset{(0.018)}{	\textrm{bipartisan basis}}$, $\underset{(0.014)}{	\textrm{good friend}}$, $\underset{(0.011)}{\textrm{bipartisan way}}$, $\underset{(0.011)}{\textrm{democratic colleagues}}$, $\underset {(0.011)}{\textrm{ common ground}} $ \\ 
   \hline
\end{tabular}
}
\caption{Most frequent terms for each topic for session 97 (left) and session 114 (right). Values in parenthesis denote appearance rates.}
\label{tab:topics}
\end{table}

%% file: tableposneg.tex
  \begin{table}[t!]
\centering
\resizebox{0.99\textwidth}{!}{%
\begin{tabular}{clcc}
 \hline
 \# &Topic& Most frequent positive terms & Most frequent negative terms \\ 
  \hline
  1& Trade & $\underset{(0.416)}{\textrm{united states}}$, $\underset{(0.04)}{\textrm{puerto rico}}$, $\underset{(0.01)}{\textrm{trade agreements}}$, $\underset{(0.009)}{\textrm{transpacific partnership}}$, $\underset{(0.007)}{\textrm{trade promotion}}$ & $\underset{(0.277)}{\textrm{united states}}$, $\underset{(0.034)}{\textrm{trade agreements}}$, $\underset{(0.021)}{\textrm{promotion authority}}$, $\underset{(0.021)}{\textrm{trade promotion}}$, $\underset{(0.015)}{\textrm{currency manipulation}}$ \\ 
  2&Export/Import & $\underset{(0.035)}{\textrm{exportimport bank}}$, $\underset{(0.03)}{\textrm{small businesses}}$, $\underset{(0.029)}{\textrm{american workers}}$, $\underset{(0.029)}{\textrm{minimum wage}}$, $\underset{(0.026)}{\textrm{exim bank}}$ & $\underset{(0.055)}{\textrm{small businesses}}$, $\underset{(0.026)}{\textrm{economic growth}}$, $\underset{(0.017)}{\textrm{american workers}}$, $\underset{(0.013)}{\textrm{exportimport bank}}$, $\underset{(0.011)}{\textrm{labor relations}}$ \\ 
  3&Commemoration & $\underset{(0.039)}{\textrm{medical research}}$, $\underset{(0.019)}{\textrm{national institutes}}$, $\underset{(0.017)}{\textrm{state university}}$, $\underset{(0.015)}{\textrm{passed away}}$, $\underset{(0.013)}{\textrm{high school}}$ & $\underset{(0.019)}{\textrm{high school}}$, $\underset{(0.018)}{\textrm{pay tribute}}$, $\underset{(0.017)}{\textrm{state university}}$, $\underset{(0.016)}{\textrm{passed away}}$, $\underset{(0.012)}{\textrm{public servant}}$ \\ 
  4&Supreme court& $\underset{(0.122)}{\textrm{supreme court}}$, $\underset{(0.034)}{\textrm{attorney general}}$, $\underset{(0.016)}{\textrm{district court}}$, $\underset{(0.015)}{\textrm{court nominee}}$, $\underset{(0.015)}{\textrm{judge garland}}$ & $\underset{(0.13)}{\textrm{supreme court}}$, $\underset{(0.028)}{\textrm{justice scalia}}$, $\underset{(0.021)}{\textrm{attorney general}}$, $\underset{(0.016)}{\textrm{presidential election}}$, $\underset{(0.011)}{\textrm{confirmation process}}$ \\ 
  5&Finances& $\underset{(0.085)}{\textrm{wall street}}$, $\underset{(0.015)}{\textrm{federal reserve}}$, $\underset{(0.012)}{\textrm{financial crisis}}$, $\underset{(0.009)}{\textrm{financial institutions}}$, $\underset{(0.008)}{\textrm{consumer protection}}$ & $\underset{(0.029)}{\textrm{wall street}}$, $\underset{(0.018)}{\textrm{federal reserve}}$, $\underset{(0.016)}{\textrm{protection act}}$, $\underset{(0.012)}{\textrm{banking housing}}$, $\underset{(0.01)}{\textrm{monetary policy}}$ \\ 
  6&Transport& $\underset{(0.075)}{\textrm{home state}}$, $\underset{(0.015)}{\textrm{important issue}}$, $\underset{(0.015)}{\textrm{critically important}}$, $\underset{(0.013)}{\textrm{transportation system}}$, $\underset{(0.011)}{\textrm{roads highways}}$ & $\underset{(0.073)}{\textrm{home state}}$, $\underset{(0.023)}{\textrm{important issue}}$, $\underset{(0.014)}{\textrm{critically important}}$, $\underset{(0.011)}{\textrm{faa reauthorization}}$, $\underset{(0.009)}{\textrm{bottom line}}$ \\ 
  7&Law enforcement& $\underset{(0.067)}{\textrm{law enforcement}}$, $\underset{(0.024)}{\textrm{gun violence}}$, $\underset{(0.022)}{\textrm{background checks}}$, $\underset{(0.021)}{\textrm{mental health}}$, $\underset{(0.021)}{\textrm{background check}}$ & $\underset{(0.078)}{\textrm{law enforcement}}$, $\underset{(0.035)}{\textrm{human trafficking}}$, $\underset{(0.017)}{\textrm{mental health}}$, $\underset{(0.016)}{\textrm{criminal justice}}$, $\underset{(0.012)}{\textrm{sex trafficking}}$ \\ 
  8&Middle East& $\underset{(0.139)}{\textrm{national security}}$, $\underset{(0.028)}{\textrm{middle east}}$, $\underset{(0.023)}{\textrm{united states}}$, $\underset{(0.013)}{\textrm{al qaeda}}$, $\underset{(0.01)}{\textrm{civil war}}$ & $\underset{(0.113)}{\textrm{national security}}$, $\underset{(0.034)}{\textrm{united states}}$, $\underset{(0.033)}{\textrm{middle east}}$, $\underset{(0.018)}{\textrm{islamic state}}$, $\underset{(0.016)}{\textrm{al qaeda}}$ \\ 
  9&Budget& $\underset{(0.02)}{\textrm{government shutdown}}$, $\underset{(0.02)}{\textrm{control act}}$, $\underset{(0.015)}{\textrm{appropriations process}}$, $\underset{(0.014)}{\textrm{budget deal}}$, $\underset{(0.012)}{\textrm{budget caps}}$ & $\underset{(0.029)}{\textrm{balanced budget}}$, $\underset{(0.022)}{\textrm{national debt}}$, $\underset{(0.017)}{\textrm{appropriations process}}$, $\underset{(0.014)}{\textrm{budget act}}$, $\underset{(0.011)}{\textrm{control act}}$ \\ 
  10&Education& $\underset{(0.058)}{\textrm{high school}}$, $\underset{(0.024)}{\textrm{left behind}}$, $\underset{(0.022)}{\textrm{higher education}}$, $\underset{(0.02)}{\textrm{child left}}$, $\underset{(0.018)}{\textrm{student loan}}$ & $\underset{(0.044)}{\textrm{high school}}$, $\underset{(0.029)}{\textrm{left behind}}$, $\underset{(0.024)}{\textrm{child left}}$, $\underset{(0.019)}{\textrm{health education}}$, $\underset{(0.019)}{\textrm{education labor}}$ \\ 
  11&Climate change& $\underset{(0.117)}{\textrm{climate change}}$, $\underset{(0.031)}{\textrm{public health}}$, $\underset{(0.029)}{\textrm{fossil fuel}}$, $\underset{(0.023)}{\textrm{carbon pollution}}$, $\underset{(0.018)}{\textrm{global warming}}$ & $\underset{(0.044)}{\textrm{public health}}$, $\underset{(0.044)}{\textrm{clean water}}$, $\underset{(0.034)}{\textrm{climate change}}$, $\underset{(0.024)}{\textrm{zika virus}}$, $\underset{(0.022)}{\textrm{water act}}$ \\ 
  12&Taxes& $\underset{(0.062)}{\textrm{middle class}}$, $\underset{(0.034)}{\textrm{republican budget}}$, $\underset{(0.023)}{\textrm{tax credit}}$, $\underset{(0.021)}{\textrm{tax breaks}}$, $\underset{(0.012)}{\textrm{tax code}}$ & $\underset{(0.036)}{\textrm{tax code}}$, $\underset{(0.035)}{\textrm{middle class}}$, $\underset{(0.03)}{\textrm{tax reform}}$, $\underset{(0.024)}{\textrm{tax credit}}$, $\underset{(0.01)}{\textrm{income tax}}$ \\ 
  13&Civil rights& $\underset{(0.04)}{\textrm{civil rights}}$, $\underset{(0.027)}{\textrm{voting rights}}$, $\underset{(0.022)}{\textrm{rights act}}$, $\underset{(0.02)}{\textrm{citizens united}}$, $\underset{(0.019)}{\textrm{equal pay}}$ & $\underset{(0.037)}{\textrm{civil rights}}$, $\underset{(0.017)}{\textrm{equal pay}}$, $\underset{(0.014)}{\textrm{fair housing}}$, $\underset{(0.012)}{\textrm{voting rights}}$, $\underset{(0.011)}{\textrm{rights act}}$ \\ 
  14&Federal government& $\underset{(0.122)}{\textrm{federal government}}$, $\underset{(0.036)}{\textrm{private sector}}$, $\underset{(0.019)}{\textrm{taxpayer dollars}}$, $\underset{(0.019)}{\textrm{federal agencies}}$, $\underset{(0.012)}{\textrm{local governments}}$ & $\underset{(0.146)}{\textrm{federal government}}$, $\underset{(0.028)}{\textrm{private sector}}$, $\underset{(0.026)}{\textrm{taxpayer dollars}}$, $\underset{(0.015)}{\textrm{federal agencies}}$, $\underset{(0.015)}{\textrm{waste fraud}}$ \\ 
  15&Health care& $\underset{(0.122)}{\textrm{health care}}$, $\underset{(0.055)}{\textrm{affordable care}}$, $\underset{(0.052)}{\textrm{care act}}$, $\underset{(0.045)}{\textrm{health insurance}}$, $\underset{(0.043)}{\textrm{planned parenthood}}$ & $\underset{(0.167)}{\textrm{health care}}$, $\underset{(0.028)}{\textrm{planned parenthood}}$, $\underset{(0.025)}{\textrm{health insurance}}$, $\underset{(0.024)}{\textrm{affordable care}}$, $\underset{(0.023)}{\textrm{care act}}$ \\ 
  16&Rhetoric& $\underset{(0.044)}{\textrm{next week}}$, $\underset{(0.02)}{\textrm{equally divided}}$, $\underset{(0.016)}{\textrm{en bloc}}$, $\underset{(0.012)}{\textrm{important legislation}}$, $\underset{(0.012)}{\textrm{final passage}}$ & $\underset{(0.062)}{\textrm{equally divided}}$, $\underset{(0.049)}{\textrm{en bloc}}$, $\underset{(0.032)}{\textrm{important legislation}}$, $\underset{(0.029)}{\textrm{next week}}$, $\underset{(0.016)}{\textrm{final passage}}$ \\ 
  17&Wars& $\underset{(0.049)}{\textrm{puerto rico}}$, $\underset{(0.047)}{\textrm{world war}}$, $\underset{(0.036)}{\textrm{war ii}}$, $\underset{(0.016)}{\textrm{vietnam war}}$, $\underset{(0.013)}{\textrm{loved ones}}$ & $\underset{(0.053)}{\textrm{world war}}$, $\underset{(0.049)}{\textrm{war ii}}$, $\underset{(0.03)}{\textrm{puerto rico}}$, $\underset{(0.017)}{\textrm{choice act}}$, $\underset{(0.013)}{\textrm{veterans administration}}$ \\ 
  18&Nuclear arms& $\underset{(0.039)}{\textrm{nuclear weapon}}$, $\underset{(0.033)}{\textrm{united states}}$, $\underset{(0.033)}{\textrm{nuclear weapons}}$, $\underset{(0.023)}{\textrm{nuclear program}}$, $\underset{(0.018)}{\textrm{irans nuclear}}$ & $\underset{(0.036)}{\textrm{united states}}$, $\underset{(0.03)}{\textrm{nuclear weapons}}$, $\underset{(0.023)}{\textrm{nuclear program}}$, $\underset{(0.02)}{\textrm{middle east}}$, $\underset{(0.02)}{\textrm{nuclear weapon}}$ \\ 
  19&Social security& $\underset{(0.115)}{\textrm{social security}}$, $\underset{(0.042)}{\textrm{trust fund}}$, $\underset{(0.022)}{\textrm{highway trust}}$, $\underset{(0.014)}{\textrm{older americans}}$, $\underset{(0.009)}{\textrm{security benefits}}$ & $\underset{(0.12)}{\textrm{social security}}$, $\underset{(0.051)}{\textrm{trust fund}}$, $\underset{(0.017)}{\textrm{highway trust}}$, $\underset{(0.012)}{\textrm{security administration}}$, $\underset{(0.009)}{\textrm{disability insurance}}$ \\ 
  20&Army& $\underset{(0.037)}{\textrm{air force}}$, $\underset{(0.036)}{\textrm{defense authorization}}$, $\underset{(0.033)}{\textrm{armed services}}$, $\underset{(0.03)}{\textrm{national defense}}$, $\underset{(0.018)}{\textrm{national guard}}$ & $\underset{(0.055)}{\textrm{air force}}$, $\underset{(0.054)}{\textrm{armed services}}$, $\underset{(0.045)}{\textrm{national defense}}$, $\underset{(0.043)}{\textrm{defense authorization}}$, $\underset{(0.024)}{\textrm{national guard}}$ \\ 
  21&Natural resources & $\underset{(0.026)}{\textrm{energy efficiency}}$, $\underset{(0.023)}{\textrm{tar sands}}$, $\underset{(0.022)}{\textrm{water conservation}}$, $\underset{(0.02)}{\textrm{conservation fund}}$, $\underset{(0.018)}{\textrm{clean energy}}$ & $\underset{(0.024)}{\textrm{keystone xl}}$, $\underset{(0.021)}{\textrm{natural gas}}$, $\underset{(0.02)}{\textrm{natural resources}}$, $\underset{(0.02)}{\textrm{xl pipeline}}$, $\underset{(0.016)}{\textrm{energy security}}$ \\ 
  22&Human rights & $\underset{(0.065)}{\textrm{human rights}}$, $\underset{(0.036)}{\textrm{state department}}$, $\underset{(0.028)}{\textrm{foreign relations}}$, $\underset{(0.02)}{\textrm{united nations}}$, $\underset{(0.017)}{\textrm{american citizens}}$ & $\underset{(0.047)}{\textrm{religious freedom}}$, $\underset{(0.037)}{\textrm{human rights}}$, $\underset{(0.033)}{\textrm{state department}}$, $\underset{(0.027)}{\textrm{foreign relations}}$, $\underset{(0.022)}{\textrm{religious liberty}}$ \\ 
  23&Security & $\underset{(0.109)}{\textrm{homeland security}}$, $\underset{(0.027)}{\textrm{cyber security}}$, $\underset{(0.019)}{\textrm{usa freedom}}$, $\underset{(0.018)}{\textrm{freedom act}}$, $\underset{(0.018)}{\textrm{immigration reform}}$ & $\underset{(0.082)}{\textrm{homeland security}}$, $\underset{(0.034)}{\textrm{cyber security}}$, $\underset{(0.025)}{\textrm{inspector general}}$, $\underset{(0.012)}{\textrm{freedom act}}$, $\underset{(0.012)}{\textrm{usa freedom}}$ \\ 
  24&Puerto Rico & $\underset{(0.013)}{\textrm{first responders}}$, $\underset{(0.013)}{\textrm{local communities}}$, $\underset{(0.011)}{\textrm{economic development}}$, $\underset{(0.011)}{\textrm{emergency funding}}$, $\underset{(0.009)}{\textrm{grant program}}$ & $\underset{(0.019)}{\textrm{puerto rico}}$, $\underset{(0.016)}{\textrm{economic development}}$, $\underset{(0.014)}{\textrm{foster care}}$, $\underset{(0.012)}{\textrm{local communities}}$, $\underset{(0.01)}{\textrm{much needed}}$ \\ 
  25&Party rhetoric & $\underset{(0.019)}{\textrm{republican colleagues}}$, $\underset{(0.015)}{\textrm{bipartisan basis}}$, $\underset{(0.012)}{\textrm{republican leadership}}$, $\underset{(0.012)}{\textrm{good friend}}$, $\underset{(0.011)}{\textrm{put together}}$ & $\underset{(0.018)}{\textrm{democratic colleagues}}$, $\underset{(0.015)}{\textrm{bipartisan basis}}$, $\underset{(0.015)}{\textrm{bipartisan way}}$, $\underset{(0.013)}{\textrm{good friend}}$, $\underset{(0.011)}{\textrm{democratic friends}}$ \\ 
   \hline
\end{tabular}
}
\caption{Most frequent positive (left) and negative (right) terms for
  each topic for session 114. Values in parenthesis denote appearance
  rates.
  \label{tab:tableposneg}}
\end{table}

%% file: tableTBIP.tex
\begin{table}[t!]
  \centering
  \resizebox{\columnwidth}{!}{%
\begin{tabular}{rrrrrrrrl}
  \hline
  
 & Min. & 1st Qu. & Median & Mean & 3rd Qu. & Max. & SD & Sessions (\#)\\ 
  \hline
  & \multicolumn{8}{c}{Most liberal Democrats}\\
Byron Dorgan (D) & 0.68 & 0.73 & 0.87 & 0.86 & 0.98 & 1.08 & 0.15 & 105$-$111 (7) \\ 
  Dale Bumpers (D) & 0.68 & 0.68 & 0.68 & 0.68 & 0.68 & 0.68 &  & 105$-$105 (1) \\ 
  Thomas Harkin (D) & 0.21 & 0.48 & 0.57 & 0.63 & 0.66 & 1.14 & 0.27 & 105$-$113 (9) \\ 
  Christopher Murphy (D) & 0.43 & 0.50 & 0.56 & 0.56 & 0.63 & 0.69 & 0.18 & 113$-$114 (2) \\ 
  Paul Wellstone (D) & 0.41 & 0.45 & 0.48 & 0.53 & 0.59 & 0.69 & 0.15
                                                              &
                                                                105$-$107 (3) \\
  \hline
  & \multicolumn{8}{c}{Most conservative Democrats}\\
  Daniel Inouye (D) & $-$0.56 & $-$0.31 & $-$0.13 & $-$0.14 & 0.08 & 0.20 & 0.28 & 105$-$112 (8) \\ 
  Robert Torricelli (D) & $-$0.27 & $-$0.20 & $-$0.13 & $-$0.16 & $-$0.10 & $-$0.06 & 0.11 & 105$-$107 (3) \\ 
  Ben Nelson (D) & $-$0.26 & $-$0.24 & $-$0.19 & $-$0.17 & $-$0.14 & $-$0.01 & 0.09 & 107$-$112 (6) \\ 
  Arlen Specter (D) & $-$0.29 & $-$0.27 & $-$0.24 & $-$0.18 & $-$0.12 & 0.00 & 0.16 & 109$-$111 (3) \\ 
  Evan Bayh (D) & $-$1.14 & $-$1.07 & $-$0.52 & $-$0.49 & 0.06 & 0.22 & 0.65 &
                                                                       106$-$111 (6) \\
  \hline
  \hline
  & \multicolumn{8}{c}{Most conservative Republicans}\\
  Jesse Helms (R) & $-$1.42 & $-$1.28 & $-$1.15 & $-$1.14 & $-$1.01 & $-$0.86 & 0.28 & 105$-$107 (3) \\ 
  Gordon Smith (R) & $-$1.49 & $-$1.34 & $-$0.77 & $-$0.80 & $-$0.27 & $-$0.14 & 0.61 & 105$-$110 (6) \\ 
  John Barrasso (R) & $-$0.94 & $-$0.89 & $-$0.67 & $-$0.71 & $-$0.57 & $-$0.47 & 0.20 & 110$-$114 (5) \\ 
  John Hoeven (R) & $-$0.91 & $-$0.77 & $-$0.62 & $-$0.69 & $-$0.59 & $-$0.55 & 0.19 & 112$-$114 (3) \\ 
  Mike Johanns (R) & $-$0.61 & $-$0.60 & $-$0.59 & $-$0.59 & $-$0.57 & $-$0.56 &
                                                                     0.03 & 111$-$113 (3) \\
  \hline
  & \multicolumn{8}{c}{Most liberal Republicans}\\
  Ed Bryant (R) & 0.01 & 0.01 & 0.01 & 0.01 & 0.01 & 0.01 &  & 106$-$106 (1) \\ 
  John Chafee (R) & 0.09 & 0.09 & 0.09 & 0.09 & 0.09 & 0.09 &  & 105$-$105 (1) \\ 
  Dirk Kempthorne (R) & 0.14 & 0.14 & 0.14 & 0.14 & 0.14 & 0.14 &  & 105$-$105 (1) \\ 
  Phil Gramm (R) & 0.06 & 0.12 & 0.18 & 0.16 & 0.22 & 0.25 & 0.10 & 105$-$107 (3) \\ 
  Howard McKeon (R) & 0.21 & 0.21 & 0.21 & 0.21 & 0.21 & 0.21 &  &
                                                                   111$-$111 (1) \\
  \hline
  \hline
  & \multicolumn{8}{c}{Independent Senators}\\
  James Jeffords (I) & $-$0.16 & $-$0.16 & $-$0.15 & $-$0.15 & $-$0.14 & $-$0.14 & 0.01 & 107$-$109 (3) \\ 
  Bernard Sanders (I) & 0.50 & 0.62 & 0.68 & 0.71 & 0.86 & 0.88 & 0.16 & 110$-$114 (5) \\ 
  Joseph Lieberman (I) & $-$0.02 & 0.01 & 0.05 & 0.05 & 0.08 & 0.12 & 0.07 & 110$-$112 (3) \\ 
  Angus King (I) & 0.09 & 0.09 & 0.10 & 0.10 & 0.11 & 0.11 & 0.02 & 113$-$114 (2) \\ 
   \hline
\end{tabular}
}
\caption{Overview on the estimated ideal points for the last ten
  Congress sessions, i.e., starting with session~105, for selected
  Democrat and Republican speakers. The five most liberal and
  conservative speakers for each party according to their average
  ideal point values are included as well as the independent
  speakers. The estimated ideal points are summarized using
  descriptive statistics (minimum -- Min.; 1st quartile -- 1st Qu.;
  Median; Mean; 3rd quartile -- 3rd Qu.; maximum - Max.; standard
  deviation -- SD). The specific sessions (Sessions) when the speaker
  was a member of the Senate during these last ten Congress sessions
  are also given together with the number of sessions in parentheses
  (\#).}
\label{tab:IPs}
\end{table}